\documentclass[9pt,preprint]{sigplanconf} %
\usepackage{url}
\usepackage{xspace} 
\usepackage{alltt}
\usepackage[T1]{fontenc}
\usepackage{graphicx}
\usepackage{amssymb}

\newcommand{\short}[1]{#1}
\newcommand{\full}[1]{}

\newcommand{\anony}[2]{#2}

\newenvironment{code}{\begin{alltt}\footnotesize}{\end{alltt}}
\renewcommand\c[1]{\mbox{\tt\footnotesize #1}} %
\newcommand\m[1]{\mbox{$#1$}} %
\newcommand{\defn}[1]{\textit{#1}} %
\renewcommand{\O}[1]{\m{O(#1)}} %
\newcommand{\F}[1]{\c{field\_#1}} %
\newcommand{\M}{\c{member}} %
\newcommand{\U}{\c{demand}} %
\newcommand{\set}[1]{\c{\{#1\}}}
\newcommand{\dom}[1]{\c{#1.keys()}} %
\newcommand{\img}[2]{\c{#1\{#2\}}} %
\newcommand{\imge}[1]{\c{#1}} %
\newcommand{\imgo}[2]{\c{#1.get(#2)}} %
\newcommand{\inv}[1]{\c{inv\_#1}} %
\newcommand{\T}[1]{\c{tag\_#1}} %
\renewcommand{\S}[1]{\c{fil\_#1}} %
\newcommand{\G}[1]{\m{G}(\c{#1})} %
\def\mathify#1{\ifmmode{\mbox{$#1$}}\else\mbox{$#1$}\fi}
\renewcommand{\=}{\mathify{\hspace{3pt}=\hspace{3pt}}\linebreak[0]}

\newcommand{\myparag}[1]{\paragraph{\rm \bf #1.}} 

\newenvironment{example}{\textbf{\textit{Example}}.}{\hfill$\blacksquare$}

\usepackage{tikz} 
\usetikzlibrary{arrows,positioning,decorations.pathreplacing,decorations.markings} 
\tikzset{
    >=stealth',
    varbox/.style={
           rectangle, rounded corners, draw=black, thick,
           text width=2.3em, minimum height=1.2em,
           text centered},
    clause/.style={
           <-, thin,
           shorten <=2pt,
           shorten >=2pt},
}
\tikzset{->-/.style={decoration={
  markings,
  mark=at position #1 with {\arrow{>}}},postaction={decorate}}
}
\tikzset{-dot-/.style={decoration={
  markings,
  mark=at position #1 with {\arrow{*}}},postaction={decorate}}
}

\newcommand{\todo}[1]{} %
\renewcommand{\omit}[1]{} %
\newcommand{\notes}[1]{} %

\begin{document}

\setlength{\pdfpageheight}{\paperheight}
\setlength{\pdfpagewidth}{\paperwidth}

\conferenceinfo{CONF 'yy}{Month d--d, 20yy, City, ST, Country} 
\copyrightyear{20yy} 
\copyrightdata{978-1-nnnn-nnnn-n/yy/mm} 
\titlebanner{}
\preprintfooter{}

\title{Demand-Driven Incremental Object Queries\vspace{-2ex}\thanks{\anony{}{
    This work was supported in part by NSF under grants 
    CCF-1414078, %
    CNS-1421893, %
    IIS-1447549, %
    CCF-1248184, %
    CCF-0964196, %
    and CCF-0613913; %
    and ONR under grants
    N000141512208 %
    and N000140910651. %
}    \vspace{-27ex} %
}}
\anony{
\authorinfo{}{\vspace{-1ex}}{\vspace{-1ex}}
}
{
\authorinfo{Yanhong A. Liu \and Jon Brandvein \and Scott D. Stoller \and Bo Lin}{Computer Science Department, Stony Brook University}{\{liu,jbrandve,stoller,bolin\}@cs.stonybrook.edu}
}
\maketitle

\begin{abstract}

Object queries are essential in information seeking and decision
making in vast areas of applications.
However, a query may involve %
complex conditions on objects and sets, which can be arbitrarily
nested and aliased.  The objects and sets involved as well as the
demand---the given parameter values of interest---can change
arbitrarily.  How to implement object queries efficiently under all
possible updates, and furthermore to provide complexity guarantees?

This paper describes an automatic method.
The method allows powerful queries to be written completely
declaratively.  It transforms demand as well as all objects and sets
into relations.
Most importantly, it defines invariants for not only the query
results, but also all auxiliary values about the objects and sets
involved, including those for propagating demand, and incrementally
maintains all of them.
Implementation and experiments with problems from a variety of
application areas, including distributed algorithms and probabilistic
queries, confirm the analyzed complexities, trade-offs, and
significant improvements over prior work.

\end{abstract}


\section{Introduction}

Consider the following query.  Given a special user, \c{celeb}, and a
group, \c{group}, as parameter values of interest, the query returns
the set of email addresses of all users who are in both the set of
followers of \c{celeb} and \c{group}, and whose location satisfies
condition \c{cond}:
\begin{code}
  // parameters: celeb, group
  \{user.email: user in celeb.followers, user in group,
               cond(user.loc)\}
\end{code}
This query can help find and monitor, for example, voters in a
political campaign, suspects in a criminal case, or targets of a
planned advertisement.
Similar queries can be about, for example, sellers of a special
product, authorized personnel of a certain organization, or
health-care providers for a particular illness, instead of followers
of a special user.

In general, such queries are essential in information seeking and decision
making, in everyday life, distributed computation, probabilistic
inference, etc.
The challenging problems are:
\begin{itemize}

\item A query can involve any number and combination of objects and
  sets, with complex conditions on them,
  and the objects and sets can be arbitrarily nested and aliased.

\item The query can be repeatedly asked 
  while the sets and objects involved starting from the given
  parameter values can change arbitrarily, %
  and the parameter values can change arbitrarily too.

\end{itemize}
While such queries can be programmed manually at a low level 
to handle all the changes efficiently,
it is much more desirable to be able to write the queries at a high
level, and have efficient implementations generated automatically.

For the given example query, 9 different kinds of updates may affect
the query result, possibly with many instances of each kind, scattered
in different places in the rest of the program.  It requires
significant effort to write efficient incremental computation code
that handles all the updates.

This paper describes an automatic method.
It consists of three main
new contributions:
\begin{enumerate}

\item It allows queries to be written completely declaratively, using
  flexible constraints on sets and objects that can be arbitrarily
  nested and aliased.

  Writing such queries arbitrarily could lead to constraints that are
  impossible to solve, including the famous Russell's paradox.
  We introduce a simple, natural
  condition %
  to exclude such cases, while ensuring that normal queries can be
  written completely declaratively.

\item It handles changes to demand, i.e., the query parameter values
  of interest, uniformly as other changes, and defines invariants for
  not only the query results, but also all auxiliary values about the
  objects and sets involved, including those for propagating demand.
  This allows the overall method to provide precise complexity
  guarantees.

  Providing complexity guarantees is extremely challenging due to
  arbitrary dynamic changes to demand and to all objects and sets that
  might be relevant. %
  With all values captured by invariants, our method uses systematic
  maintenance of all invariants to support precise complexity
  calculation.

\item It generates standalone efficient incremental maintenance code
  that handles arbitrary changes to arbitrarily nested and aliased
  sets and objects\notes{} without resorting to additional runtime support.

  This is done by transforming everything into flat relations, similar
  to prior work, %
  but
  then generating complete, properly ordered maintenance code to tie
  the maintenance of all invariants together, with appropriate tests
  to ensure correct maintenance under nesting and aliasing.

\end{enumerate}

We have developed IncOQ, a prototype implementation of the method, and used
it to experiment with complex queries from a variety of applications,
including those from the most relevant previous
work~\cite{Liu+06ImplCRBAC-PEPM,Willis06,RotLiu08OSQ-GPCE,Willis08,Gor+12Compose-PEPM}
and from new applications in distributed
algorithms~\cite{Liu+12DistPL-OOPSLA,Liu+12DistSpec-SSS} and probabilistic
queries~\cite{milch2007,milch2010extending,blog16examp-github}.  Our
evaluations consider all important factors: asymptotic time and space
complexities, constant-factor optimizations, demand set size, query-update
ratio, auxiliary indices, runtime overhead, demand propagation strategies,
and transformation time and other characteristics.

There is a large amount of related work on object queries, incremental
computation, and demand-driven computation, as discussed in
Section~\ref{sec-relate}.  Previous works do not
support fully declarative object queries with a simple well-defined
semantics; they handle limited queries and updates or require
sophisticated runtime support to handle demand and dynamic updates
and are less efficient;
and they do not provide complexity guarantees for such complex queries
and updates.  They also do not evaluate the wide %
variety of important factors as we do. %

\notes{}%

\full{}

\full{}

\section{Language and problem description}
\label{sec-lang}

Our method applies to any language that supports the following object
query and update constructs.

\notes{}%

\notes{}%

\myparag{Object queries}

Object queries are queries over objects.
Precisely, an \defn{object query} is a comprehension of the following
form plus a set of \defn{parameters}---variables %
whose values are bound before the query:
\[\small
\begin{array}{@{}r@{~}c@{~}l@{}}
query      & ::= & 
     \verb|{|\,result~\verb|:|~\,(membership \,|\, condition\,)^+ \verb|}| \\
membership & ::= & variable \verb| in | selector \\
selector   & ::= & variable \,|\, selector.field \\
condition  & ::= & expression\\%~~\c{\scriptsize//\,any side-effect-free expression}\\
result     & ::= & expression\\%~~\c{\scriptsize//\,any side-effect-free expression}\\
\end{array}
\]
where \m{expression} is any expression that is function of the values
of \m{variables} and \m{selectors} in the expression.

For a query to be well-formed, we require that every variable in the
query be \defn{reachable} from a parameter, i.e., be a parameter or
recursively be the left-side variable of a membership clause whose
variable on the right side %
is reachable.

Objects, including set objects, have reference semantics.  
Object equality is reference equality. %
Given values of parameters, a query returns the set of values of the
result expression for all combinations of values of variables that
satisfy all membership and condition clauses.
\full{}\short{If} an error occurs during the evaluation
of a condition or the result expression, that combination of values
is skipped.
A formal semantics is given in the context of an object-oriented
language\anony{~\cite{Liu+14DistPL-arxiv-anony}}{ that includes also
  constructs for distributed programming~\cite{Liu+14DistPL-arxiv}}.

We can see that an object query %
may contain arbitrary object field selections and set membership
constraints, where objects and sets can be arbitrarily nested and
aliased.
Our method just requires that each condition and the result expression
be a function of 
the variables and selectors in the query.
This gives us the freedom to decide when to evaluate a condition or
the result expression.

The queries are similar to those defined by Rothamel and
Liu~\cite{RotLiu08OSQ-GPCE}, but there, membership and condition
clauses are ordered, as \c{\m{membership^+ condition^*}}, albeit
superficially, and the requirements specified are not completely
correct---due to an insufficient requirement on variables without
using reachability and an overly strong requirement on conditions and
the result expression being functions of only variables in the query.

\myparag{Flexible constraints on sets and objects}

\notes{} Note that we allow all membership and condition
clauses to be written in any order, so each clause is simply a
constraint, and the entire query is completely declarative.  

It is well known that arbitrary constraints could lead to queries that
have no meaningful answers, including Russell's paradox:
\begin{itemize}
  \setlength{\itemsep}{-.75ex}
\item[] If \c{s} is the set \c{\{x: not x in x\}}, is \c{s} in \c{s}?
\item[] If \c{s} is in \c{s}, then by definition of \c{s}, \c{s}
  should not be in \c{s}.
\item[] If \c{s} is not in \c{s}, then by definition of \c{s}, \c{s}
  should be in \c{s}.
\end{itemize}
Well-known solutions, led by language SETL, require the use of a
membership clause to enumerate a variable first, and later uses of the
variable are simply tests.  So, \c{\{x: not x in x\}} is not allowed.

However, this ordering forces queries to be less declarative.
\full{} %
For example, the order of the four clauses in the following query in our
language does not affect the semantics or efficiency.
\begin{code}
    \{y-x: x in s, y in t, y > x, x < 5\}
\end{code}
Consider the following three out of 24 (=4!) possible orders of the
four clauses written in Python syntax, or other similar syntax in
SETL, Haskell, etc.
In Python, the second query below may be much faster than the first,
whereas the third gives a runtime error.  
In Haskell, the third gives a collection of unevaluated subtractions
instead of the results of the subtractions.
In SETL, the last two are not legal, because all bindings of
variables must go first.
\begin{code}
    \{y-x for x in s for y in t if y > x if x < 5\}
    \{y-x for x in s if x < 5 for y in t if y > x\}
    \{y-x for x in s if y > x for y in t if x < 5\}
\end{code}
Even database and logic languages typically have requirements on the
order of conditions, e.g., in SQL, variables must be bound in the
\c{from} clause before used in the \c{where} clause.  In some logic
languages that do relax such requirements, it is well known that
non-stratified negation may arise, with no commonly agreed upon
semantics, especially in cases that encode paradox queries like
\c{\{x: not x in x\}} using recursive rules.

Our requirement for queries to be well-formed removes the ordering
requirement to allow queries to be completely declarative, while
excluding abnormal cases.  For example, \c{\{x: not x in x\}} is
well-formed if \c{x} is a parameter, 
and not well-formed otherwise.
Our transformation method determines a best implementation for
satisfying all the constraints while efficiently handling all possible
changes.

Note that our queries are not recursive, whereas queries in logic
languages may be, but logic languages do not have arbitrarily nested
and aliased sets and objects as our language has.

\myparag{Updates to objects}

Our method handles all possible updates to the values that the query
depends on.
Note that the result of an object query may depend on not only the
values of query parameters, but also other values---specifically, all
objects reachable by following the object fields and set elements used
starting from the values of the parameters.
There are three kinds of fundamental updates to the values that a
query depends on, besides assignments to query parameters; our method
decomposes all updates into combinations of these fundamental updates,
and handles all combinations of these fundamental updates: %
\begin{itemize}
  \setlength{\itemsep}{-.5ex}
\item \c{o.f = x}---assign value \c{x} to field \c{f} of object \c{o}
\item \c{s.add(x)}---add element \c{x} to set \c{s}
\item \c{s.del(x)}---delete element \c{x} from set \c{s}
\end{itemize}%
\notes{}

\myparag{Cost}
Object queries are expensive if evaluated straightforwardly\full{}.  Let
\c{s1}, ..., \c{sk} be largest possible sets of values obtained from
evaluating the right sides of membership clauses in a query, and {\small
  \m{time(conditions)}} and {\small \m{time(result)}} be the times for
evaluation of the conditions and the result expression, respectively.  Let
\c{\#s} denote
the number of elements in set \c{s}.  The cost of
straightforward evaluation of the query is
\full{}
\short{{\small$O(\c{\#s1} \times \ldots \times \c{\#sk} \times (time(conditions) + time(result)))$}.}

The fundamental updates each take \O{1} time.

\myparag{Running example}

\full{}
\notes{}%
We use the example query in Section~1 as a running example.
We indicate the parameters in comments for clarity; they are bound
variables before the query, determined automatically.

The query is well-formed because all variables in the
query---\c{celeb}, \c{group}, and \c{user}---are reachable from the
parameters.
The query satisfies the requirement about condition and result
expressions if \c{cond(user.loc)} is a function of \c{user} and
\c{user.loc}---for example, by testing \c{user.loc} against, say,
\c{"NYC"} or some range of geographic coordinates.  We assume that the
requirement is satisfied and the test takes \O{1} time.

If the query is run straightforwardly, by iterating using each
membership clause, it takes \O{\c{\#celeb.followers}\times\c{\#group}}
time; even if run optimally, it takes %
\O{\c{\#celeb.followers}+\c{\#group}} time, because the sets used must
at least be read and processed.

There are 9 possible kinds of fundamental updates that may affect the query
result: 2 for change of demand by assigning to parameters \c{celeb}
and \c{group},
4 for adding to and deleting from sets \c{celeb.followers} and
\c{group}, and 3 for assigning to fields \c{celeb.followers},
\c{user.loc}, and \c{user.email}.

\myparag{Demand-driven incremental computation}

Not only are object queries complex and expensive, they can also be
repeated while the sets and objects involved change.  Therefore, it is
important to be able to compute the query results incrementally with
respect to the changes.
Furthermore, query parameters may take on any combination of values,
and it may be impossible to determine the values statically.
Therefore, it is essential to be able to incrementally maintain the
query result for parameter values that are determined dynamically on
demand.

\full{}

\full{}%
\short{ 
We define a set of \defn{demand parameters} of a query to be a
subset of the parameters such that the query is still well-formed if only
the demand parameters are given as parameters.

\begin{example}
  For the running example, both \c{celeb} and \c{group} must be demand
  parameters for the query to be well-formed, but if a new clause,
  \c{celeb in group}, is added in the query,
  then the query is still well-formed if only \c{group} is taken as a
  demand parameter.
\end{example}

Note that fewer demand parameters mean more parameter values for which
the query results may be maintained.

\begin{example}
  For the running example, if \c{celeb in group} is added and
  \c{group} is the only demand parameter, then the query result may be
  maintained for all \c{celeb} values in \c{group} instead of just the
  given parameter value of \c{celeb}.
\end{example}

We define the \defn{demand set}, \U, of the query to be a set of
combinations of values of the demand parameters.  }

\full{} 

\short{%
  \begin{itemize}
  \item 
  The program can explicitly specify demand parameters, and add and
  delete elements to and from \U{} to control what query results are
  to be maintained, or specify a replacement strategy, such as the
  least-recently-used, to use when space is short.

  \item

   Alternatively, by default, our method considers all query
   parameters as demand parameters,
  adds the values of these\notes{} parameters to \U{}
  when these values are first queried on, and deletes them when it can
  be determined conservatively that
  these values will never be queried on again.
  \end{itemize}
}

Given queries, updates, and demands, the problem is to incrementally
maintain the query results, at the updates, for all values of demand
parameters in \U.  Our method guarantees that
each execution of a transformed query expression returns the same
value as the original query expression.

\full{}

We consider \U{} to be relatively small\notes{}, compared with the set of all possible
combinations of values of all query parameters.\full{}

\notes{}

\section{Method and notation}

Our method for transforming a single query has three phases.
\begin{description}

\item {\bf Phase 1} transforms all object queries, demands, and
  updates into relational queries and updates. 

  It transforms demands, as well as objects and sets that can be
  arbitrarily nested and aliased, all into flat relations that are not
  aliased.  This allows queries over arbitrarily nested and aliased
  objects and sets for any demand set to be incrementalized in a
  simpler and uniform way.

\item {\bf Phase 2} generates efficient implementations for relational
  queries and updates by exploiting constraints\notes{} from the objects, updates, and demands.

  The key idea is to define and incrementally maintain invariants for
  not only the query results, but also all auxiliary values about the
  objects and sets involved, including those for propagating demand.
  This allows systematic maintenance of invariants to take all the
  objects, demands, and updates into account in generating efficient
  implementations.

  This phase then generates complete, properly ordered maintenance
  code to tie the maintenance of all invariants together, with
  appropriate tests to ensure correct maintenance under nesting and
  aliasing.

\item {\bf Phase 3} transforms the implementations on relations back
  to implementations on objects.

  This makes best use of given objects and sets in the original
  program as well as maps for auxiliary values to minimize the time
  and space of the resulting program.

\end{description}
\full{}

\myparag{Overall algorithm}
Our overall algorithm handles any number of queries, including nested
queries and aggregate queries.  
Handling multiple queries, including
nested queries, is done by repeated application of the method for
handling a single query.
As an optimization, to avoid transforming between the object and
relational domains multiple times, our method does Phase 1 for all
queries, then Phase 2 for all queries, and then Phase 3 for all
queries.

For independent queries, i.e., queries whose results do not depend on
the results of other queries, Phase 2 can be done for them in any
order.  For a dependent query, i.e., a query whose result depends on
the results of other queries, including subqueries in a nested query,
Phase 2 is done following the chain of dependencies---first for those
other queries and then for the dependent query.
\short{Aggregate queries employ a library of rules specialized to the
  aggregate operations \c{count}, \c{max}, etc.
}

Details %
of support for nested queries, aggregate queries, and other features for
ease of efficient queries can be found in~\cite{Bra16thesis}.

\myparag{Relational queries}
Relational queries are queries over flat relations, i.e., sets of flat
tuples.  Precisely, a \defn{relational query} has the same form as an
object query except that in membership clauses, the left side can also
be a tuple form, not just a variable, and the right side can only be a
variable, not a field selection, 
\full{}\short{i.e.,}
\[\small
\begin{array}{@{}r@{~}c@{~}l@{}}
membership  & ::= & \verb|(|variable^+\verb|) in | variable \\
\end{array}
\]
\full{}
Relational queries are just SQL queries but expressed using a simpler
syntax and where tuple components are referred to by position numbers
instead of names\full{}.

Relational queries are also expensive if evaluated straightforwardly\full{}.
Our method uses constraints from
objects, %
demands, and updates
to help optimize incremental relational queries.

\myparag{Generated code using operations on relations}
Our generated code for efficient relational queries uses the following
operations on relations, besides usual operations on sets:
\begin{itemize}

\item \c{R add x} and \c{R del x}, \defn{counted addition and
    deletion}, increments and decrements, respectively, the count for
  the number of times \c{x} has been added to but not deleted from
  \c{R}, and keeps \c{x} in \c{R} iff its count is at least 1.

  These correspond to a way of implementing bag element addition and
  deletion, contrasting standard set element addition and deletion,
  done using assignments \c{R = R+\{x\}} and \c{R = R-\{x\}},
  respectively, where \c{+} and \c{-} denote set union and difference,
  respectively.
  
\item \img{R.(j1,...,jh)}{(i1,...,ik)=(x1,...,xk)}, \defn{image set
    of\linebreak \c{(x1,...,xk)} under \c{R}, mapping components
    \c{(i1,...,ik)} to components \c{(j1,...,jh)}}, is the set of
  values of components \c{j1}, ..., \c{jh} of tuples of \c{R} whose
  components \c{i1}, ..., \c{ik} have values \c{x1}, ..., \c{xk},
  respectively.

  \begin{example}
    If \c{R} is a relation of arity at least 4, i.e., a set of tuples
    of at least 4 components, then \img{R.(4,1)}{(1,3)=(x,y)} is the
    set of values of components 4 and 1 of tuples in \c{R} whose
    components 1 and 3 equal the values of \c{x} and \c{y},
    respectively.
  \end{example}

  If \c{i1}, ..., \c{ik} is\notes{} a
  prefix of the list of indices, we can omit
  \c{(i1,...,ik)=}\,;\full{}\short{ and}
  if \c{j1}, ..., \c{jh} is the list of remaining indices, we can omit \c{.(j1,...,jh)}.\full{}
  If a tuple has only one component, we can omit \c{()}.\full{}
  If the tuples in \set{} are empty, we can omit \set{()=()}.

  \begin{example}
    If \c{R} is a relation of arity 4,
    \img{R.(3,4)}{(1,2)=(x,y)} can be abbreviated as 
    \img{R.(3,4)}{(x,y)}, 
    \img{R}{(1,2)=(x,y)}, and
    \img{R}{(x,y)}.

    \img{R.(4)}{(2)=(x)} can be abbreviated as \img{R.4}{2=x}---the
    set of values of component 4 of tuples in \c{R} whose component 2
    equals the value of \c{x}.

    \img{R.4}{()=()} can be abbreviated as \c{R.4}---the set of values
    of component 4 of the tuples in \c{R}.
  \end{example}

\item When \c{R} is a binary relation, \inv{R}, \defn{inverse relation
    of \c{R}}, is the set of pairs in \c{R} with the first and second
  components switched, i.e., \inv{R} \= \set{(y,x):~(x,y) in R}.

  Thus, \img{\inv{R}}{x} \= \img{R.1}{2=x} whereas \img{R}{x} \=
  \img{R.2}{1=x}.

\end{itemize}
We use a map to implement the mapping from values of components
\c{i1}, ..., \c{ik} to values of components \c{j1}, ..., \c{jh} for
\c{R}.  This uses a nested structure, like a trie, with one level for
each component \c{i1}, ..., \c{ik}; hashing is used to implement the
sets at each level.  This allows the image set operation to take
expected \O{\c{\#image}} time, where \c{image} is the resulting image
set, to enumerate the elements, or \O{1} time, to return just a
reference to the resulting set.  The map is updated in expected \O{1}
time for each element addition to and deletion from \c{R}.  The space
taken by the map is \O{\c{\#R}}.

\myparag{Generated code using operations on objects}
Our method generates code in a conventional object-oriented
programming language that supports sets, maps, and tuples and where
all values are objects.  Besides set element addition and deletion and
object field assignment, operations in the following table
are used.
Sets and maps are empty when first created.
Tuples\notes{} are of constant
length and are immutable.
Our method adds membership tests to guard the given updates---\c{x not
  in s} to guard \c{s.add(x)}, and \c{x in s} to guard
\c{s.del(x)}---when the test results cannot be determined statically,
so that updates do not propagate unnecessarily.
\begin{center}
\footnotesize
\begin{tabular}{@{\,}l@{~}|@{~}l@{~}}
\hline
\c{s.cadd(x)}   & add element \c{x} to counted set \c{s} \\
\c{s.cdel(x)}   & delete element \c{x} from counted set \c{s} \\
\c{x in s}      & return whether \c{x} is an element of set \c{s} \\
\hline
\c{m.add(x,y)}  & add \c{y} to the image set of key \c{x} under map \c{m}\\
\c{m.del(x,y)}  & delete \c{y} from the image set of key \c{x} under map \c{m}\\
\c{m.cadd(x,y)} & add \c{y} to the counted image set of key \c{x}
                  under map \c{m} \\
\c{m.cdel(x,y)} & delete \c{y} from the counted image set of key \c{x}
                under map \c{m} \\
\c{\dom{m}}     & return the set of keys in map \c{m} \\
\c{\imgo{m}{x}} & return the value of key \c{x} under map \c{m} \\
\hline
\c{(x1,...,xk)} & create a tuple with components \c{x1}, ..., \c{xk} \\
\hline
\c{x isset}      & return whether \c{x} is a set \\
\c{x hasfield y} & return whether \c{x} has field \c{y} \\
\hline
\end{tabular}
\end{center}
Each of these operations takes expected \O{1} time\notes{}\notes{}.

We use standard statements for assignment (\c{v = e}), sequencing
(\c{stmt1 stmt2}), branching (\c{if b:~stmt}), and looping (\c{for v
  in s:} \c{stmt}).
We abbreviate assignment \c{v = v op e}, where \c{op} is any binary
operation, as \c{v op= e}.
\full{}
We assume that all bound variables in the program are renamed so they
are distinct.

\section{Phase 1: Transform 
  into relational queries and updates}
\label{sec-rel}

Phase 1 transforms each object query and its demand parameters into a
relational query, and transforms updates as well.  
Transformations of queries and updates are as in a prior
work~\cite{RotLiu08OSQ-GPCE}; only the simple addition of demand is
new, but it will be used substantially in Phase 2 to define auxiliary
relations to contain only objects that are reachable from values in
the demand set.

\myparag{Transform object queries into relational queries}

\full{}
\short{We use the following relations.
For each field \c{f}, relation \F{f} relates an object with the
  value of the field \c{f} of the object; that is,
  (o,x) \m{\in} \F{f} \m{\iff} x \m{=} o.f.
  Relation \M{} relates each set with each member of the set; that
  is,
  (s,x) \m{\in} \M{} \m{\iff} x \m{\in} s.
}

To transform an object query into a relational query, the following
two rules are applied repeatedly until they do not apply:
\begin{itemize}

\item For each variable \c{o} and field \c{f}, 
  replace 
  all occurrences of the field selection \c{o.f} with a fresh
  variable, say \c{x}, and add a new membership clause \c{(o,x) in
    \F{f}}.
  
\item Replace each membership clause \c{x in s}, where \c{x} and \c{s}
  are variables, with a new membership clause \c{(s,x) in \M}.

\end{itemize}
So, for example, a sequence of field selections are transformed by applying
the first rule repeatedly from left to right.

\full{}
\todo{}

\begin{example}
In the running example, this yields the following, where 
\c{e}, \c{fs}, and \c{l} are fresh variables:
\begin{code}
  // parameters: celeb, group
  \{e: (user,e) in \F{email},
      (celeb,fs) in \F{followers}, (fs,user) in \M,
      (group,user) in \M, (user,l) in \F{loc},
      cond(l)\}
\end{code}\vspace{-3.25ex}
\end{example}

\myparag{Add demands to relational queries}

We transform the demands into an additional relational query
constraint, by adding a membership clause, \c{(dp1,...,dpj) in \U}, in
the query, constraining the values of demand parameters 
\c{dp1,...,dpj} to be in \U.

\begin{example}
For the running example, with demand parameters \c{celeb} and
\c{group}, %
this adds \c{(celeb, group) in \U}, yielding:
\begin{code}
  // parameters: celeb, group
  \{e: (celeb,group) in \U, (user,e) in \F{email},
      (celeb,fs) in \F{followers}, (fs,user) in \M,
      (group,user) in \M, (user,l) in \F{loc},
      cond(l)\}
\end{code}\vspace{-3.25ex}
\end{example}

\myparag{Transform updates to objects into updates to relations}

Updates to objects, including set objects, are transformed into
updates to the field relations and member relation.
\begin{itemize}
  \setlength{\itemsep}{0ex}
\item \c{o.f = x} is transformed into the two updates below, or only
  the second update if \c{o.f} had no value before:\vspace{-1ex}
\begin{code}
  \F{f} -= \{(o,o.f)\}
  \F{f} += \{(o,x)\}
\end{code}\vspace{-1ex}
\item \c{s.add(x)} is transformed into \c{\M{} += \{(s,x)\}}.
\item \c{s.del(x)} is transformed into \c{\M{} -= \{(s,x)\}}.
\end{itemize}

\begin{example}
For the running example, the transformed updates are additions to and
deletions from \M, \F{email},\linebreak \F{followers}, and \F{loc}.
\end{example}

\section{Phase 2: Incrementalize under updates with filtering by demands}
\label{sec-inc-fil}

Phase 2 incrementalizes the relational query from Phase 1 with respect
to updates and demands.  We present it in two steps to show how to
minimize both the running time and space usage: Step 2-INC generates
efficient incremental maintenance with respect to the updates, and
Step 2-FIL extends Step 2-INC to generate efficient incremental
maintenance filtered with demands.  Both steps provide precise cost
guarantees.

\subsection{Generate incremental maintenance}
\label{sec-inc}

Step 2-INC first stores the query result in a fresh relation \c{r},
with components for
parameters \c{p1}, ..., \c{pk} and the result of the query, and
maintains an invariant of the form
\begin{code}
   r \= \{(p1,...,pk, \m{result}): \m{(\,membership\,|\,condition\,)\sp{+}}\}
\end{code}
Thus, \img{r}{(p1,...,pk)} equals the query result, and the query is
replaced with the\short{ constant-time} retrieval
\img{r}{(p1,...,pk)}\full{}, a reference to
the result set.
\todo{}

The result set %
is updated as the values it depends on are updated.  Where necessary,
a copy of the result set is made and used,
at cost linear in the size of the set.  Techniques exist for
determining where copying is necessary~\cite{Freudenberger:83,GoyPai98}.

\begin{example}
For the running example, the following invariant is maintained:\vspace{-1ex}
\begin{code}
r\=\{(celeb,group,e)
    : (celeb,group) in \U, (user,e) in \F{email},
      (celeb,fs) in \F{followers}, (fs,user) in \M,
      (group,user) in \M, (user,l) in \F{loc},
      cond(l)\}
\end{code}\vspace{-1ex}
and the query is replaced with \img{r}{(celeb,group)}.
\end{example}

Step 2-INC then generates efficient incremental maintenance code for
updates that correspond to each membership clause in the right side of
the invariant, for example, for updates \c{\F{loc} +=} \c{\{(user,l)\}}
and \c{\F{loc} -= \{(user,l)\}} that correspond to the last clause.
The key challenge is to find an optimal order of accessing variables
through relations in the other clauses to arrive at needed updates to
the query results, corresponding to the well-known join order problem.

The basic ideas for generating incremental maintenance code are as in
prior work,
e.g.,~\cite{PaiKoe82,Liu+06ImplCRBAC-PEPM,RotLiu08OSQ-GPCE}.  The main
new ideas here to address the key challenge are (1) formulate the
problem %
as an optimal growing edge cover problem, with the cost for each
growth step captured precisely and symbolically, and (2) use the
constraint from demands, together with constraints from objects and
updates as in a prior work~\cite{RotLiu08OSQ-GPCE}, as good heuristics
for solving the problem, despite its worse-case factorial time, as
follows.\vspace{-.8ex}
\begin{itemize}
  \setlength{\itemsep}{-.4ex}
\item[(1)] Create a query graph: take variables as vertices,
  and clauses as directed edges (directed hyperedges in the case of \U)
  labeled with the relation used in the clause, i.e.,
  take clause \c{(x,y) in R} as directed edge \c{(x,y)} labeled \c{R}, and
  take clause \c{(dp1,...,dpj)\hspace{-.5ex} in\hspace{-.2ex} \U} as directed hyperedge
  \c{(dp1,...,dpj)} labeled \U;
  different occurrences of the same relation are suffixed with an integer
  indicating their order of occurrence. %

\item[-] Repeatedly select an edge to access its source and target vertices
  (its different subsequences of vertices in the case of a hyperedge)
  until all edges are covered.

\item[-]
  Analyze the cost of selecting each next edge, \c{(x,y)} labeled
  \c{R}, to access its source and target vertices (with \c{x} and
  \c{y} denoting two different subsequences of vertices in the case of
  a hyperedge, and with the corresponding sequences of indices
  used in place of 1 and 2 below):\\
  -- \O{1} if both \c{x} and \c{y} have been accessed already,\\
  -- \O{\c{\#\img{R.2}{1=x}}} %
  if \c{x} has been accessed but \c{y} has not,\\
  -- \O{\c{\#\img{R.1}{2=y}}}
  if \c{y} has been accessed but \c{x} has not,\\
  -- \O{\c{\#R}} if neither \c{x} nor \c{y} have been accessed before.

\item[-]
  For \c{k} clauses, there are \O{\c{k}!}  possible orders to consider
  in the worse case.

\item[(2)] Use the heuristic of selecting minimum-cost edges first to
  drastically reduce the search space, exploiting \O{1} or small cost
  edges based on constraints from objects, updates, and demand:\\
  -- \c{(o,x) in \F{f}} when \c{o}  has been accessed, for any f, \\
  -- \c{(x,y) in R} corresponding to the update being handled, \\
  -- \c{(dp1,...,dpj) in \U}.

  The first two are \O{1}, and the last one is %
  relatively small.
\end{itemize}

\begin{example}
For the running example, the relational query has the following
  query graph:\vspace{-4ex}
\begin{center}
\begin{tikzpicture}[node distance=4ex, auto, scale=0.86, transform shape]
 \node[] (dummy) {};
 \node[varbox, inner sep=5pt, above=.75cm of dummy] (Celeb)
   {\c{celeb}};
 \node[varbox, inner sep=5pt, right=1.5cm of Celeb] (Group)
   {\c{group}}
   edge[clause] node[anchor=west] {
     \begin{tabular}{@{~}c@{}} \hspace{-5ex}\c{\U}\\~ \end{tabular}} (Celeb);
 \node[varbox, inner sep=5pt, left=0cm of dummy] (Fs)
   {\c{fs}}
   edge[clause] node[anchor=east] {\c{\F{followers}}~} (Celeb);
 \node[varbox, inner sep=5pt, below=.75cm of dummy] (User)
   {\c{user}}
   edge[clause] node[anchor=east] {\c{\M\_1}~} (Fs)
   edge[clause, bend right=0] node[anchor=west] {~\c{\M\_2}} (Group);
 \node[varbox, below=1.6cm of Fs ] (L)
   {\c{l}}
   edge[clause] node[anchor=east] {\c{\F{loc}}~~} (User);
 \node[varbox, right=.5cm of L] (E)
   {\c{e}}
   edge[clause] node[anchor=west] {~\c{\F{email}}} (User);

 \draw[-dot-=.55] (Celeb) to (Group);
 \draw[-dot-=.55] (Celeb) to (Fs);
 \draw[-dot-=.55] (Fs) to (User);
 \draw[-dot-=.55] (Group) to (User);
 \draw[-dot-=.55] (User) to (L);
 \draw[-dot-=.55] (User) to (E);

\end{tikzpicture}\\%\vspace{-108ex}\\\nopagebreak
\end{center}
For updates \c{\F{loc} +=} \c{\{(user,l)\}} and \c{\F{loc} -=
  \{(user,l)\}}, out of 5! possible orders, only two are considered
following the heuristics in (2) above, shown in
Table~\ref{tab-covers-small}.
\begin{table}[htb]
  \centering
  \small
\begin{tabular}{@{\,}l@{}l@{\,}|@{\,}l@{\,}}
  \hline
 &edges followed                       & cost factors\\
  \hline\hline
 &\c{(user,e)} labeled \F{email}      & \c{1}\\
 &\c{(group,user)} labeled \c{\M\_2}  & \c{\#\img{\inv{\M}}{user}}\\
 &\c{(celeb,group)} labeled \U        & \c{\#\img{\U.1}{2=group}}\\
 &\c{(celeb,fs)} labeled \F{followers}& \c{1} \\
 &\c{(fs,user)} labeled \c{\M\_1}     & \c{1} \\
  \hline
 &\c{(user,e)} labeled \F{email}      & \c{1}\\
 &\c{(fs,user)} labeled \c{\M\_1}     & \c{\#\img{\inv{\M}}{user}}\\
 &\c{(celeb,fs)} labeled \F{followers}& \c{\#\img{\inv{\F{followers}}}{fs}}\\
 &\c{(celeb,group)} labeled \U        & \c{\#\img{\U.2}{1=celeb}}\\
 &\c{(group,user)} labeled \c{\M\_2}  & \c{1}\\
\end{tabular}
  \caption{Growing edge covers for update to \F{loc}}
  \label{tab-covers-small}
\end{table}\\
If no other constraints are given, the first order has the minimum
cost.  If it is given that there is only one \c{celeb} in\linebreak
\img{\inv{\F{followers}}}{fs} and only one \c{group} for each
\c{celeb} in \U, then the second order has the minimum cost.
\end{example}

Afterward, code generation uses the following basic algorithm:
\begin{enumerate}
  \setlength{\itemsep}{.25ex}

\item Each ordering obtained gives an order of generated incremental
  maintenance clauses---\c{for} clauses for %
  unbound variables and \c{if} clauses for %
  bound variables, as in~\cite{Liu+06ImplCRBAC-PEPM,RotLiu08OSQ-GPCE}.

\item A condition clause can be inserted anywhere after all variables
  used in it are bound, but least expensive clauses are inserted
  earliest.

\item A final statement updates
  the query result relation \c{r}.

\end{enumerate}

\begin{example}
  For the running example, the first order gives the following
  incremental maintenance code for update \c{\F{loc} +=} \c{\{(user,l)\}}.
  Comments show asymptotic costs.
\begin{code}
if cond(l): 
  for e in \img{\F{email}}{user}:         // 1
    for group in \img{\inv{\M}}{user}:    // \#\img{\inv{\M}}{user}
      for celeb in \img{\U.1}{2=group}: // \#\img{\U.1}{2=group}
        for fs in \img{\F{followers}}{celeb}:  // 1
          if (fs,user) in \M:          // 1
            r add (celeb,group,e)
\end{code}
Because \inv{\M} is used, it is also maintained incrementally.  
For \c{\M{} += \{(group,user)\}}, the following incremental
maintenance is generated:
\begin{code}
   \inv{\M} += \{(user,group)\}
\end{code}%
If \inv{\F{loc}} is used in\full{} incremental maintenance of any query,
then, for \c{\F{loc} += \{(user,l)\}}, the following incremental
maintenance is also generated:
\begin{code}
   \inv{\F{loc}} += \{(l,user)\}
\end{code}\vspace{-3ex}
\end{example}

Note that, for addition to \U, which binds the query parameters to a
new combination of values, incremental maintenance considers matching
values of all other clauses, which corresponds to computing from
scratch, as can be expected.
For updates to \F{f} and \M, incremental maintenance considers
only matching values for the remaining clauses led to by the added or
deleted tuple, and thus can be much more efficient when \U{} is small.

\notes{}

\subsection{Generate filtered incremental maintenance}
\label{sec-fil}

Step 2-FIL uses \U{} to more significantly reduce running times and
space usage, by filtering the auxiliary relations
to contain only tuples that are ``reachable'' from values in \U.
Intuitively, this is done by ``tagging'' objects following the graph
edges starting from objects in \U.

Formally, this is the key idea of defining and maintaining invariants,
for not only the query results, but also all auxiliary values about
the objects and sets involved, including those for propagating demand.

\begin{example}
  For the running example, in the generated code in
  Section~\ref{sec-inc},
  the for-loop that binds \c{group} iterates over the set of all
  groups that the updated \c{user} is in, but we can filter that set
  to contain only the groups in
  \U.  Also, the generated code need not be run at all if the updated
  \c{user} is not in the follower set of any user in \U{} or any group
  in \U.
\end{example}

\myparag{Define invariants for filtered maintenance}
The algorithm for defining invariants for capturing demand has two
steps:
\begin{enumerate}

\item For each variable \c{var} (such as \c{group} in the above
  example) whose value is retrieved through an image set operation, or is a
  variable (such as \c{user} in the above example) in the clause for
  the update being handled, find a maximal directed acyclic subgraph
  \G{var}, of the query graph, that reaches the vertex for \c{var}
  starting from vertices for the variables in \U.

\item Define a \defn{tag set} for each vertex, and a \defn{filtered
    relation} for each edge, in the subgraph \G{var}, mutually
  recursively:
  \begin{itemize}
    \setlength{\itemsep}{-.4ex}

  \item [--] The \defn{tag set for a vertex} \c{v} is the intersection of
    objects projected from the filtered relations for the incoming
    edges \c{e1}, ..., \c{ek} of \c{v}:
\begin{code}
tag_v\=\{v: (u1,v) in fil_e1,\,...,\,(uk,v) in fil_ek\}
\end{code}

  \item [--] The \defn{filtered relation for an edge} \c{e} is the
    relation filtered using objects in the tag set for the source vertex
    of \c{e}:
\begin{code}
fil_e\=\{(u,v): (u,v) in e, u in tag_u\}
\end{code}

We define filtered relations also for inverse member and inverse fields
used to retrieve values in tag sets.

  \item [--] Initially, \defn{the tag set for a vertex for a demand
    parameter} \c{dpi}, \c{i\=1}, ..., \c{j}, is projected from
    \c{demand}:
\begin{code}
tag_dpi\=\{dpi: (dp1,...,dpj) in demand\} 
\end{code}

  \end{itemize}

  These definitions form the additional invariants that capture the
  demand precisely.
\end{enumerate}
Incremental maintenance of these invariants automatically realizes
incremental propagation of demand.

To obtain filtered maintenance, in the incremental maintenance code,
replace uses of the original relations with the corresponding filtered
relations, and add tests to check that the objects in the updates are in
the corresponding tag sets.

\begin{example}
  Continue the running example.  For the use of\linebreak \img{\inv{\M}}{user}
  to retrieve the value of \c{group}, %
  a single-vertex subgraph, containing only the vertex for \c{group},
  is found.

  Then, the following tag set and filtered relation are defined for
  \c{group}, and the original use of \inv{\M} is replaced with
  \c{\S{\inv{\M}}\_2}; suffix \c{\_2} distinguishes this occurrence of
  \inv{\M} from other occurrences. %
\begin{code}
   \T{group} \= \{group: (celeb,group) in \U\}
   \S{\inv{\M}}_2 \= \{(user,group)
                        : (user,group) in \inv{\M},
                          group in \T{group}\}
\end{code}

These auxiliary values are then incrementally maintained, just as for
any query.  This yields the following maintenance code for \c{\U{} +=
  \{(celeb,group)\}}, following the dependency chain:
\begin{code}
   // maintain \T{group} for \c{\U{} += \{(celeb,group)\}}
   \T{group} add group\smallskip
   // maintain \S{\inv{\M}\_2} for \c{\T{group} add group}
   for user in \img{\M}{group}:
     \S{\inv{\M}\_2} += \{(user,group)\}
\end{code}
and the following maintenance for \c{\inv{\M} += \{(user,group)\}},
i.e., for \c{\M{} += \{(group,user)\}}:
\begin{code}
   if group in \T{group}: 
     \S{\inv{\M}\_2} += \{(user,group)\}
\end{code}

Similarly, tag sets and filtered relations are defined and maintained
for \c{user}, and the original maintenance code is preceded with the test
\c{user in \T{user}}.
Test to check that \c{l} in the update \c{\F{loc} += \{(user,l)\}} is in
\c{\T{l}} is not needed, because it is implied by \c{user in \T{user}}.

We obtain the following filtered incremental maintenance code; the two
comments indicate the changes from the maintenance code in
Section~\ref{sec-inc}.
\begin{code}
if user in \T{user}:                       // added
  if cond(l): 
    for e in \img{\F{email}}{user}:
      for group in \img{\S{\inv{\M}\_2}}{user}: // use filtered
        for celeb in \img{\U.1}{2=group}:
          for fs in \img{\F{followers}}{celeb}:
            if (fs,user) in \M:
              r add (celeb,group,e)
\end{code}
\vspace{-3ex}
\end{example}

This filtering based on demand can reduce both time and auxiliary
space significantly, because \U{} is small compared to the entire data
set. %
\begin{itemize}

\item In the cost formulas for time, sizes of the relations used are
  replaced with sizes of the corresponding filtered relations.  The
  resulting formulas are used to determine an optimal order of
  retrieving matching values.

\item The auxiliary space used is also reduced, to be proportional to
  only elements of relations reachable from parameter values in \U,
  instead of %
  all objects.

\end{itemize}

\begin{example}
  Filtering reduces the cost of maintenance for \c{\F{loc}} \c{+=
    \{(user,l)\}}
from $O(\c{\#\img{\inv{\M}}{user}}\times \c{\#\img{\U.1}{2=group}})$ 
to $O(\c{\#\img{\S{\inv{\M}}\_2}{user}}\times \c{\#\img{\U.1}{2=group}})$,
which is \O{1} when \c{\#\U{}} is \O{1}, and reduces the auxiliary
space from %
number of members of all groups to only members of the groups in \U.
Relation \inv{\M} is\full{} not used in incremental
maintenance---it is dead and eliminated.
\end{example}

Note that filtered relations for member and fields, unlike for inverse
member and inverse fields, could be omitted to save the space for storing
them; the original member and field relations, together with the tag sets
for the source vertices, can be used instead.  This does not affect the
asymptotic time complexities.

\myparag{Organize all maintenance}
To organize all maintenance for a query, to ensure that all invariants
hold, even under arbitrary nesting and aliasing, four rules are used
to properly order the maintenance blocks and
add appropriate tests.
\begin{enumerate}
\item %
  Maintenance of the query result and of all auxiliary values can be
  placed %
  before or after the update to a relation.

  This is because the maintenance does not use the relation being
  updated %
  but the element being added or deleted.

  We place the maintenance after an addition update and before a
  deletion update, for overall simpler ordering.

\item For an element addition update, the maintenance of the query
  result is placed after the maintenance of the auxiliary values used.

  This is because the element being added may join with elements to be
  added to filtered auxiliary values to yield elements to be added to
  the query result.

\item For an element deletion update, the maintenance of the query
  result is placed before the maintenance of the auxiliary values
  used.

  This is because the element being deleted may join with elements to
  be deleted from filtered auxiliary values to yield elements to be
  deleted from the query result.

\item If a relation \c{R} occurs in multiple membership clauses (same
  \F{f} or \c{\M}) in the query, 
  we augment the clause %
  for each occurrence of \c{R} before the one for which we are
  generating a maintenance block,
  with a test to exclude the element being added or deleted.

  This ensures that an update that may affect a query result through
  multiple membership clauses, under nesting or aliasing, has the
  correct overall effect---that is, contributes to the maintenance of
  the query result as a single update---after combining separate
  maintenance blocks for updates corresponding to each of the clauses.

  An example is given in Appendix~\ref{sec-self-join}.

\end{enumerate}

\full{}
\vspace{-.5ex}

\myparag{Complexity guarantees}

Overall, our method provides a precise asymptotic complexity bound on
the running time of the generated code as follows: for nested loops,
multiply the cost associated with each \c{for} clause, i.e., the
number of elements to iterate over, and the sum of the cost of
evaluating other conditions and the result expression;
for concatenated blocks, sum the cost of each block.

The space complexity for the auxiliary values used is bounded
asymptotically by the size of the given data, because tag sets are
subsets of the given sets and objects, and filtered relations are
subsets of the given set membership and object field relations.  More
precisely, this is bounded by only the number 
of elements of relations %
reachable from parameter values in \U, as opposed to the size of all 
given data.

\full{} %

\section{Phase 3: Transform back to implementations on objects}
\label{sec-back}

\full{}

Phase 3 transforms operations on relations in all maintenance code and
query result retrieval back to operations on objects\full{}.
Transformations for \F{f}, \M{}, and query results are as in a prior
work~\cite{RotLiu08OSQ-GPCE}; transformations for \U{}, tag sets, and
filtered relations are new.

In each case, the operation is transformed to best use given object
fields and sets in the original program as well as auxiliary maps for
efficiency, and to add appropriate tests to ensure correctness in any
context.
\short{%
  Detailed transformations are given in Appendix~\ref{sec-phase3}; we
  show the results here through the running example.}
\full{}%

\begin{example}
  For the running example, this yields, from the filtered maintenance
  code in Section~\ref{sec-fil}, the following
  maintenance code for \c{\F{loc} += \{(user,l)\}}, i.e., for update
  \c{user.loc = l} where \c{user.loc} had no value before:
\begin{code}
if user in \T{user}:   // line 1 of maint code in Sec 5.2\todo{}
  if cond(l):          // line 2
    if user hasfield email:          // line 3
      e = user.email                 //
      if user in \dom{\S{\inv{\M}}_2}:         // line 4
        for group in \imgo{\S{\inv{\M}}_2}{user}:  //
          if group in \dom{\U{}21}:         // line 5
            for celeb in \imgo{\U{}21}{group}:  //
              if celeb hasfield followers:  // line 6
                fs = celeb.followers        // 
                if fs isset:             // line 7
                  if user in fs:         //
                    r.cadd((celeb,group), e)  // line 8
\end{code}
and maintenance of auxiliary values at updates to \U{} and \M{} as in
Section~\ref{sec-fil}, except for using \c{cadd} for \c{add}.

The retrieval of the query result is transformed into\linebreak
\imgo{r}{(celeb,group)}.
\end{example}

\myparag{Constant-factor optimizations}
Unnecessary use of counted sets can be eliminated, as discussed below, and
type information and static analysis~\cite{Gor+10Alias-DLS} can help remove
the tests that use \c{hasfield}, \c{isset}, and \c{keys}, to give
constant-factor improvements.

A main constant-factor optimization is counting elimination.  Result
relations in general need to be implemented using counted sets, i.e.,
having a count of how many times each element is added to the set.
Doing this for all maintained results, including all auxiliary values,
is an obvious overhead.  We remove counts in several main cases:
\begin{itemize}

\item When all possible updates to a set are additions.

\item When there are deletions, but there is only one possible
  combination of values of variables in the query for each element in
  the result set, 
  based on two conditions:

  (1) each variable in the query is a parameter or a variable in the
  \m{result} expression, or has its value determined uniquely by such
  a variable by following \F{f} edges from the variable; and

  (2) the \m{result} expression returns different values for different
  combinations of values of variables in it, e.g., it is a variable or
  a tuple of variables.

\item When it is any filtered relation.

\end{itemize}
\notes{}

\full{} %

\full{} %

\full{}

\full{}

\full{}
\full{}

\section{Experimental evaluation}
\label{sec-expe}

We implemented our method in a prototype system,\anony{}{ IncOQ,}
available \anony{in a public source code repository~\cite{incoq-github}}{at
  \url{http://github.com/IncOQ/}},
and performed a large number of experiments to evaluate the method and
confirm the analyzed complexities, advantages, and trade-offs.

Our system takes programs written in Python as input,
transforms queries and updates in the program according to our method,
and outputs the resulting program.
For queries in the input program, the system interprets all clauses of
the form \c{for x in s} and \c{if x in s} in the queries as membership
clauses, and the rest as condition clauses.
\full{}

\full{}

As part of the evaluation, we also experimented with two best
available previous implementations of incremental object queries: Java
Query Language (JQL)~\cite{Willis06,Willis08} and Object-Set Queries
(OSQ)~\cite{RotLiu08OSQ-GPCE}, including
porting OSQ from Python 2.5 to Python 3.4 for better comparison.

\full{}

Except for experiments using JQL, all experiments were run using
Python 3.4.2,
under 64-bit Windows~8.1 on an Intel Core~i5 4200M CPU at 2.50~GHz
with 8GB memory.
Each measured execution
of a program occurred in its own process, with the garbage collector
disabled.
\full{} All reported times are CPU time, in seconds,
  as the mean of repeated runs until the standard deviation is less
  than 10\% of the mean, with at least 10 runs. %\todo{}
\full{} %
Additional space is the size
  of the additional sets and maps used;
  the size of a set is the number of elements, and the size of a map
  is the number of keys plus the size of the image set of each key.
  \full{}
\full{}

All JQL experiments used JQL version 0.3.3, with Java 1.6.0\_45,
AspectJ 1.7.2, and ANTLR 3.0.1 and were run on the same machine as
running Python.  The CPU times reported are the mean of
50 repeated runs, as in~\cite{Willis08}.

\subsection{Asymptotic performance, demand, and constant factors ---
  via the running example query}

We use the running example query to evaluate %
asymptotic performance, the effect of demand set size, and
constant-factor optimizations, because the analyzed complexities and
generated code\notes{} have already been
discussed in detail.
Similar results were observed for other queries too.

\myparag{Asymptotic time and space performance}

We describe experiments using uniformly distributed random data
because they are simpler and easier to understand, even tiny
differences such as described in Footnote 1.  We also experimented
with Zipf distribution, well known for data studied in physical and
social sciences, and observed similar asymptotic performance
improvement. %

Table~\ref{tab-asymp} summarizes the analyzed time complexities for
the running example query and update to a user location, for the
realistic case of one user in \img{\inv{\F{followers}}}{fs} for each
\c{fs} (i.e., each \c{followers} set \c{fs} belongs to one user) and
a constant number of groups in \U{}. %
The update times for the incremental program and incremental program
with filtering (called ``filtered program'' for brevity) correspond to
the second order in Table~\ref{tab-covers-small},
because the condition for it to have the minimum cost holds for the
realistic case.
The implemented heuristic chooses this order automatically.
\begin{table}[htb]
  \centering
  \vspace{-1ex}
  \small
\begin{tabular}{@{~}l@{\,}|@{~}l@{~}|@{~}l@{\,}}
  \hline
  program  & query & update \\\hline\hline

  \begin{tabular}[t]{@{}l@{}}
    original\\
    (Section\,\ref{sec-lang})
  \end{tabular}
  &
  \begin{array}[t]{@{}l@{}}
    \m{O}(\c{\#celeb.followers}\\
    ~~~~\,\m{+}\, \c{\#group})
  \end{array}
  & \O{1}
  \\\hline

  \begin{tabular}[t]{@{}l@{}}
    incremental\\
    (Section\,\ref{sec-inc})
  \end{tabular}
  & \O{1}
  &
  \begin{array}[t]{@{}l@{}}
    \m{O}(\c{\#\img{\inv{\M}}{user}}\\
    ~~~~\,\m{\times}\, \c{\#\img{\U.2}{1=celeb}})
  \end{array}
  \\\hline

  \begin{tabular}[t]{@{}l@{}}
    filtered\\
    (Section\,\ref{sec-fil})
  \end{tabular}
  & \O{1}
  &
  \begin{array}[t]{@{}l@{}}
    \m{O}(\c{\#\img{\S{\inv{\M}}\_1}{user}}\\
    ~~~~\,\m{\times}\, \c{\#\img{\U.2}{1=celeb}})
  \end{array}
\end{tabular}
\caption{Analyzed time complexities.%
}
\label{tab-asymp}
\end{table}

The analyzed additional space is \O{\c{\#\inv{\M}}} 
for the incremental program and \O{\c{\#\S{\inv{\M}}\_1}} 
for the filtered program.

We use test data with up to 20,000 users and 1\% as many groups.  Each
user follows 0.5\% of all users, is in 5\% of the groups, and is at
one of 20\full{} locations, one of which satisfies the
location condition in the query.  There are 3 special users and 1
group (i.e., 3 pairs, where each user is paired with the same group)
in \U, where each special user has 0.5\% of all users as followers
(more followers only make the original query take linearly longer to
run, not making the updates or other queries take longer).
Thus,\full{} in the time complexity formula, \c{\#celeb.followers},
  \c{\#group}, and \c{\#\img{\inv{\M}}{user}} are linear in the number
  of users, and \c{\#\img{\S{\inv{\M}}\_1}{user}} and
  \c{\#\img{\U.2}{1=celeb}} are constants;\full{} in the space complexity formula, \c{\#\inv{\M}} is quadratic in
  the number of users, and \c{\#\S{\inv{\M}}} is linear.\full{}

\begin{figure}[htb]
  \centering
  \vspace{-2.5ex}

  \hspace{-2ex}
  \includegraphics[width=3in]{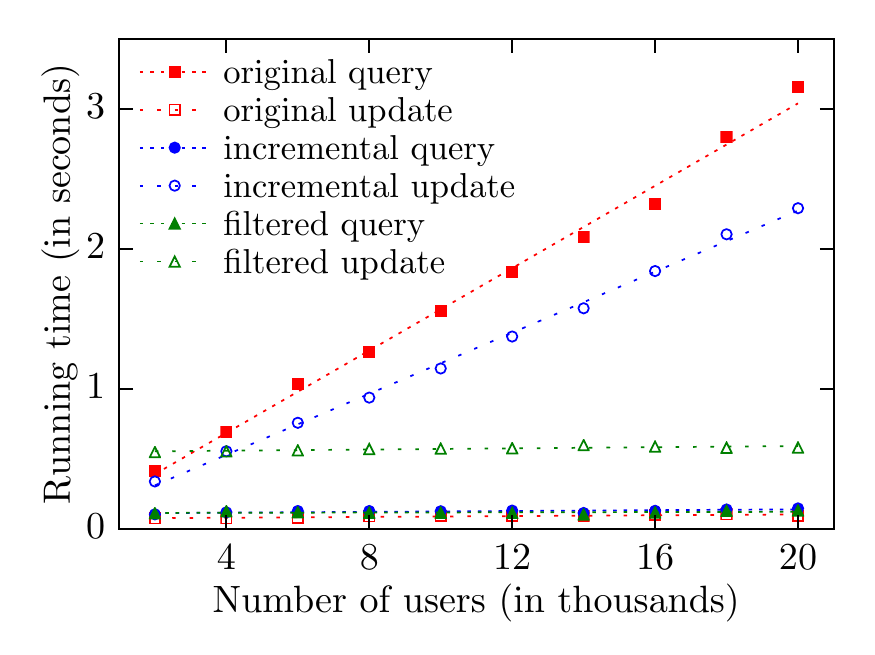}\vspace{-2.5ex}

  \hspace{-2ex}
  \includegraphics[width=3in]{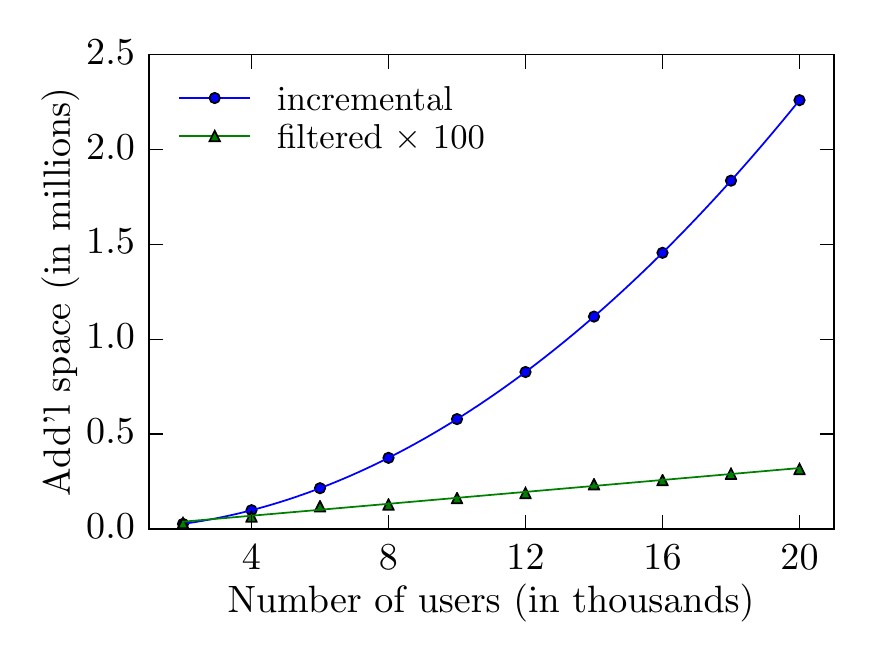}\vspace{-2.5ex}

\caption{Running time and additional space for varying input size.}
\label{fig-celeb-asymp}
\end{figure}

Figure~\ref{fig-celeb-asymp}-top shows the measured query and update
times for 200,000 queries and 200,000 updates, respectively, obtained
by measuring (1) the total time of 200,000 repeats of a query-update
pair, querying on a random user-group pair in \U{} and updating the
location of a random user, and (2) the total time without the queries,
and subtracting (2) from (1).
Figure~\ref{fig-celeb-asymp}-bottom shows the additional space
measured at the end of (1).
It confirms the analyzed %
complexities:
\begin{itemize}
  \setlength{\itemsep}{0ex}
\item The query time grows linearly with the number of users
  for the original program, and remains constant for the incremental
  and filtered programs.
\item The update time is constant for the original program, grows
  linearly for the incremental program, and is constant albeit a
  larger constant for the filtered program.\footnote{There is actually
    a tiny increase for the original and filtered.  It (and a tiny
    portion of the increase for the incremental) is due to increased
    cache misses in randomly accessing increasingly more users for
    location updates.  We confirmed this by running experiments to
    measure cache misses.  }
\item The additional space %
  is quadratic in the number of users for the incremental program, and
  linear for the filtered program.
\end{itemize}

\myparag{Effect of demand set size}
\short{%
  We also evaluated the effect of demand set size on performance, as
  described in Appendix~\ref{sec-demand-set-size}.  This helped
  confirm that filtering using demand provides significant gains when
  the demand set is small relative to the entire domain, but not
  otherwise.
} %

\myparag{Constant-factor optimizations}
\short{%
  We further measured the benefit of %
  eliminating unnecessary use of counted sets and of optimizations
  enabled by alias and type analysis, %
  as described
  in~\anony{\cite{BraLiu16OverheadElim-PEPM16-anony}}{\cite{BraLiu16OverheadElim-PEPM16}}.
  We found that the former is generally significant, while the latter
  is relatively small.}

\full{}

\full{} %

\subsection{Effect of query-update ratio, auxiliary indices, and other
  implementation factors --- via JQL queries}
\label{sec-expe-jql}

We evaluated how query-update ratio affects the performance of our
generated programs, using all three query benchmarks tested for
incrementalization of object queries implemented in
JQL~\cite{Willis08}.%
\footnote{JQL evaluation~\cite{Willis08} also studies performance
  benefit on Robocode, a Java application.
  Our work studies queries in Python applications in
  Section~\ref{sec-expe-more}.}
Their three queries, in our syntax, are
\begin{code}
   // parameters: attends
   \{a: a in attends, a.course == COMP101\}  \smallskip
   // parameters: attends, students
   \{(a, s): a in attends, s in students, 
        a.course == COMP101, a.student == s\}  \smallskip
   // parameters: attends, students, courses
   \{(a, s, c): a in attends, s in students, 
        c in courses, a.course == COMP101,
        a.student == s, a.course == c\}
\end{code}
We discovered, unexpectedly, that the JQL implementation, even though
in Java, is significantly slower than our generated Python programs,
and even asymptotically slower for join queries.

We use the same setup as described in~\cite{Willis08} for JQL
experiments: 1000 objects in each source collection, performing 5000
operations, each being either a query or a random addition and removal
of an element of \c{attends}.

\begin{figure}[htb]
  \centering
  \vspace{-2ex}

  \hspace{-2ex}
  \includegraphics[width=3in]{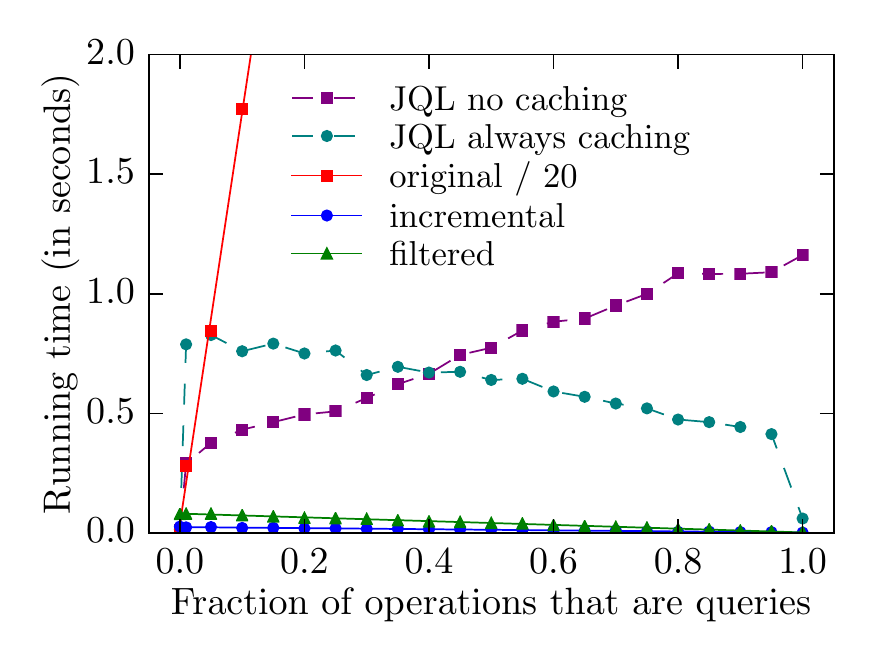}\vspace{-2.5ex}
  \caption{Running times in comparison with JQL on JQL benchmark 2 for
    varying query-update ratio.}
  \label{fig-jql-ratio}
\end{figure}

Figure~\ref{fig-jql-ratio} shows the running times of the benchmark
for the second query as the ratio of queries over updates increases:
(1) the time of original Python program increases, similar to JQL with
no caching, and
(2) the time of incremental and filtered Python programs decrease,
similar to JQL with always caching.
The crossover point depends on the query and the implementation, but
the incremental Python program outperforms all other programs
significantly---it outperforms filtered here because there is only one
fixed (reference) value for each query parameter (\c{attends}, etc.)
in \U, and all objects in \c{attends} are in the tag set for
\c{attends}, making the maintenance of tag sets and filtered relations
unnecessary overhead.\full{}

Figure~\ref{fig-jql-ratio} also shows, unexpectedly, that
(1) the incremental and filtered Python programs are faster than JQL
with incremental caching, even though Python is\full{}
much slower than Java, and
(2) these Python programs appear even asymptotically faster.
There is also a larger time advantage of our generated Python programs
over JQL with caching for the third query, and smaller for the first.
By examining the implementation of JQL, we believe that (1) the
runtime overhead of AspectJ used to implement JQL contributed to a
constant-factor slowdown, consistent with runtime overhead of dynamic
methods in general, such as OSQ discussed in
Section~\ref{sec-expe-osq}, and (2) JQL does an
inefficient\notes{} join with each additional source collection,
yielding asymptotic slowdowns compared to our method of\short{
  using}\full{} efficient
auxiliary\short{ indices}\full{}.
To confirm the asymptotic time difference, we modified the benchmarks
to not vary the query-update ratio, but to increase the size of the
source collections.

\begin{figure}[htb]
  \centering
  \vspace{-2ex}

  \hspace{-2ex}
  \includegraphics[width=3in]{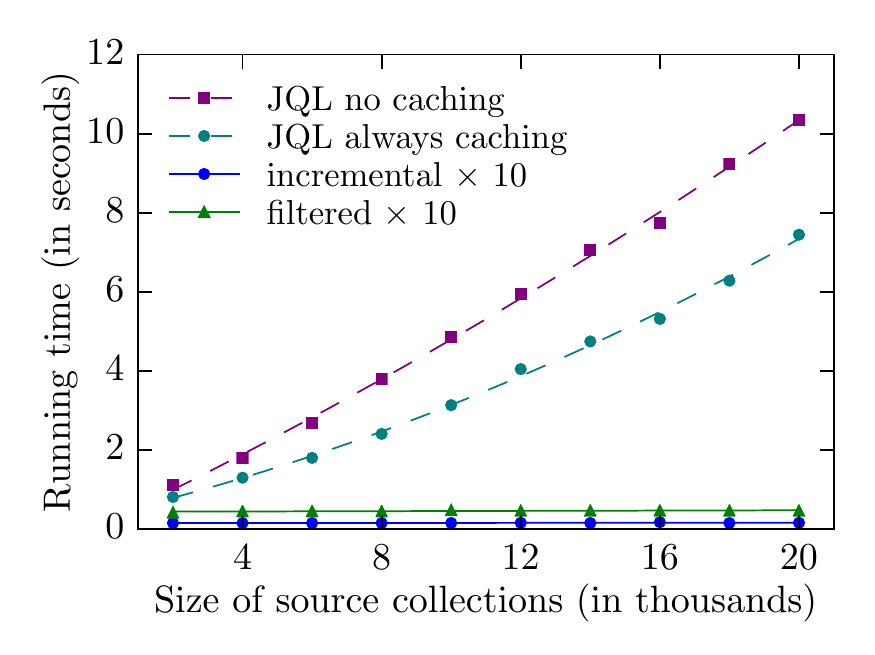}\vspace{-2.5ex}

  \caption{Running times in comparison with JQL on JQL benchmark 2 for
    varying input size (times of the original program was too large
    and thus omitted).}
  \label{fig-jql-asymp}
\end{figure}

Figure~\ref{fig-jql-asymp} shows the running times of the modified
benchmark for the second query, for collections of sizes 2,000 to
20,000, for equal numbers of queries and updates.
It shows a continued %
increase %
by JQL, both with no caching and with always caching, whereas running
times of both our generated programs stay constant.
For the third query, the increase by JQL is even larger,
whereas the times of both our programs again stay constant.  For the
first query, JQL programs are about 2-5 times slower than ours.
We also confirmed that enabling garbage collection has little effect
on the running times of our generated
programs, %
with only an average variation of about 7\%.

\subsection{Runtime overhead and demand propagation strategies --- via
  OSQ queries}
\label{sec-expe-osq}

We evaluated savings of runtime overhead by our method compared to a
dynamic method, by comparing with OSQ for incremental object queries
in Python~\cite{RotLiu08OSQ-GPCE}.  OSQ generates similar code for
maintaining the query result but uses dynamic assignment of
obligations to objects to track demand and invoke the query result
maintenance code, instead of using tag sets and filtered relations.
We found that such a dynamic method is not only less efficient, but
also error-prone, and hard to understand and optimize.  \full{}\short{The} method in this paper can improve over
OSQ asymptotically.

We implemented all the benchmarks of OSQ %
in~\cite{RotLiu08OSQ-GPCE} and compared the running times of our
filtered programs and OSQ.
We observed that our method produces the same asymptotic speedups as
OSQ, but our filtered programs are consistently faster\full{}.
For example, for the Django authorization query below, our filtered
program is over 20\% faster; other improvements measured are even
larger, to over 50\%.
\begin{code}
   // parameters: users, uid
   \{p.name: u in users, g in u.groups, p in g.perms,
        u.id == uid, g.active\}
\end{code}

More importantly, we determined that OSQ has a less refined strategy
than our method for propagating demand for queries that involve
intersection,
and that strategy can yield asymptotically worse performance.
For example, for our running example, OSQ uses only the first
membership clause to filter using demand, not both clauses as our
method does.
However, we could not demonstrate the performance difference using the
OSQ implementation on our running example, due to a bug we discovered
in OSQ; we also found OSQ difficult to understand and improve.
Instead, because our invariant-based method and implementation allow
us to easily define and implement different strategies for filtering
using demand, we implemented the OSQ strategy by defining it using
invariants, and generated the corresponding filtered programs using
our system.

We use the same setup as for Figure~\ref{fig-celeb-asymp} except with
each user in 5 groups and with all users in \U.
Using our demand propagation strategy, only users in the single group
in \U{} are in the tag set for \c{user}, and so maintenance is needed
only for updates to these users, which total 500 on average (1\% as
many groups as users, with 5 groups per user).  Using OSQ's strategy,
all users following any user in \U{} are in the tag set for \c{user},
and so maintenance
is run for updates to all such users.

\begin{figure}[htb]
  \centering
  \vspace{-2.5ex}

  \hspace{-2ex}
  \includegraphics[width=3in]{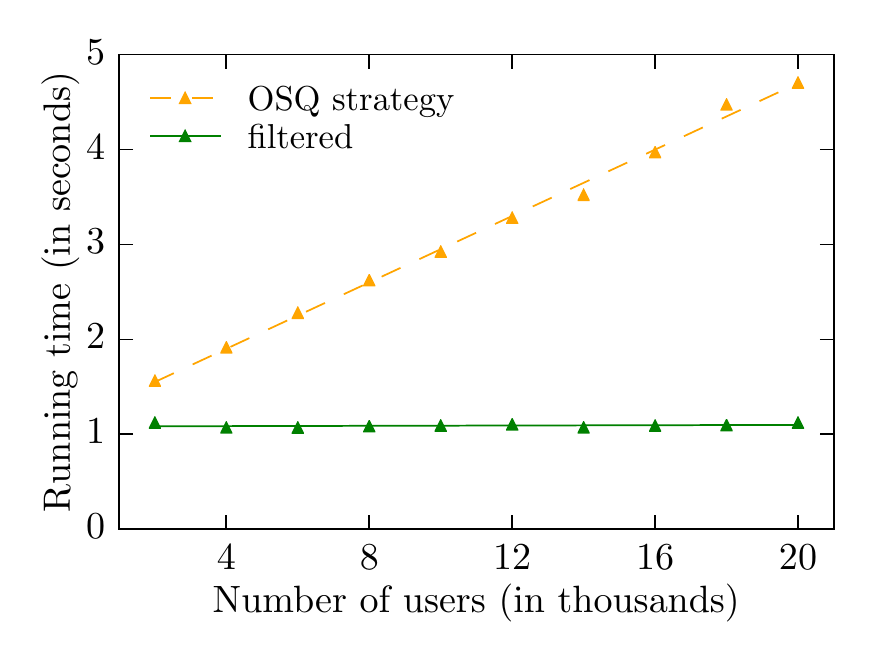}\vspace{-2.5ex}

  \caption{Running times in comparison with using OSQ's demand
    propagation strategy.}
  \label{fig-osq}

\end{figure}

Figure~\ref{fig-osq} shows the running times of our filtered program
in comparison with the filtered program that uses OSQ's demand
propagation strategy.
It confirms that our method improves over OSQ asymptotically:
the running time increases linearly for OSQ,
but is constant for our generated program. %

\subsection{Transformation time and other %
measures --- via access %
control, distributed algorithms, and probabilistic queries}
\label{sec-expe-more}

\todo{}

We also applied our system to other examples, including the
\c{CheckAccess} query and all 16 queries in Core
RBAC~\cite{Liu+06ImplCRBAC-PEPM}, the \c{SSD} and \c{DSD} constraints
in Constrained RBAC~\cite{Gor+12Compose-PEPM}, and
the most difficult
queries in a set of distributed algorithms~\cite{Liu+12DistPL-OOPSLA}
and in approximate probabilistic
inference~\cite{blog16examp-github,blog16examp-gsite}.
We summarize the most interesting results here.

Our filtered programs for RBAC automatically allow RBAC to be run with
multiple instances, where arbitrary data can be passed around
dynamically among the instances, and all operations of all instances
are run incrementally.  This was not possible
previously~\cite{Liu+06ImplCRBAC-PEPM}.
The overhead of\full{} the filtered program when there is only
one instance is %
small compared to\full{}\short{ our} incremental program
and the incremental program from~\cite{Liu+06ImplCRBAC-PEPM}, and in
fact the tag sets and filtered relations help reduce incremental
maintenance time.  For example, for the \c{CheckAccess} query in Core
RBAC, the time of incremental maintenance for creating and deleting
sessions is reduced by about 67\%.\todo{}

Our filtered programs for queries in distributed algorithms are larger
and somewhat\todo{} slower than those generated using a
previous system~\cite{Liu+12DistPL-OOPSLA},
mostly due to the generality of the new implementation, compared to
the specialized, tediously manually written incrementalization rules
used in~\cite{Liu+12DistPL-OOPSLA}; we have not focused on reducing
code size, because it does not affect the running time as much.  For
example, for Lamport's distributed mutual exclusion, the filtered
program is 480 lines compared to 124\notes{} before,
and it uses more CPU time than the incrementalized program
from~\cite{Liu+12DistPL-OOPSLA} by about 45\%\todo{} for 50
processes and less\todo{} for more processes, but it still gains
asymptotically over the original program.

Our incremental and filtered programs for queries in probabilistic
inference are all constant time under every possible update that may
be used in Markov chain Monte Carlo (MCMC)
sampling~\cite{milch2010extending}, whereas the original queries take
linear or quadratic time in the domain size.

Table~\ref{tab-all} contains a list of programs for which we have done
extensive evaluations to confirm analyzed performance and trade-offs.
The original programs are small, at most 174 lines, but the
incremental and filtered programs grow significantly, up to 2645
lines.  Manually writing incremental and filtered programs is
challenging, not only because of their size, but much more because of
complex interactions among all demands and updates, typically
scattered in many places.
Every program where 1 to 8 queries are incrementalized is transformed in
about a half minute or less, except that transforming Core RBAC and two
distributed algorithms with 16 to 22 queries each takes longer but less
than two and a half minutes;
this is consistent with the transformation time being linear in
program size.\todo{}
\todo{}

\begin{table*}[htb]
  \small
  \centering
  \vspace{-3ex}
\begin{tabular}{@{\,}l@{\,}|@{~}l@{\,}
    ||@{\,}r@{\,}|@{\,}r@{\,}|@{\,}r@{\,}
    ||@{\,}r@{\,}|@{\,}r@{\,}
    ||@{\,}r|@{\,}r@{\,}}
  \hline
  name & description
  &
  \begin{tabular}[t]{@{}c@{}} original \\ (LOC)
  \end{tabular}
  &
  \begin{tabular}[t]{@{}c@{}} query \\ (count)
  \end{tabular}
  & 
  \begin{tabular}[t]{@{}c@{}} update \\ (count)
  \end{tabular}
  & 
  \begin{tabular}[t]{@{}c@{}} inc.\ \\ (LOC)
  \end{tabular}
  &
  \begin{tabular}[t]{@{}c@{}} inc.\ trans.\ \\ time (s)
  \end{tabular}
  &
  \begin{tabular}[t]{@{}c@{}} filtered \\ (LOC)
  \end{tabular}
  &
  \begin{tabular}[t]{@{}c@{}} fil.\ trans.\ \\ time (s)
  \end{tabular}
  \\\hline\hline

  Running & running example query
  & 40 &   1 &   9 &     251 & 10.12 &     491 &  21.73 \\\hline
  \hline

  JQLbench1 & JQL benchmark for query 1~\cite{Willis08}
  & 41 &   1 &   3 &     101 &  2.02 &     136 &   3.50 \\\hline

  JQLbench2 & JQL benchmark for query 2~\cite{Willis08}
  & 43 &   1 &   5 &     143 &  4.45 &     204 &   7.23 \\\hline

  JQLbench3 & JQL benchmark for query 3~\cite{Willis08}
  & 46 &   1 &   6 &     162 &  5.97 &     249 &  10.41 \\\hline
  \hline

  WiFi & WiFi access point query~\cite{Rot08thesis} %
  & 28 &   1 &   6 &     217 & 10.95 &     369 &  19.16 \\\hline

  Auth & Django authorization query~\cite{RotLiu08OSQ-GPCE}
  & 34 &   1 &   8 &     232 & 18.72 &     468 &  33.92 \\\hline
  \hline

  Access & \c{CheckAccess} query in Core RBAC~\cite{Liu+06ImplCRBAC-PEPM}
  &139 &   3 &  13 &     322 &  6.50 &     546 &  14.00 \\\hline

  CoreRBAC & all queries in Core RBAC~\cite{Liu+06ImplCRBAC-PEPM}
  &139 &  16 &  21 &    1074 & 39.83 &    2645 & 141.31 \\\hline

  SSD & SSD constraint in Constrained RBAC~\cite{Gor+12Compose-PEPM}
  & 29 &   3 &   8 &     295 &  7.20 &     348 &   8.89 \\\hline
  \hline

  La mutex & Lamport's distributed mutual exclusion~\cite{Lam78}
  & 44 &   6 &   6 &     403 &  9.31 &     480 &  11.50 \\\hline
  RA mutex & Ricart-Agrawala's distributed mutual exclusion~\cite{RA81}
  & 55 &   8 &   4 &     683 & 18.03 &     804 &  23.11 \\\hline

  RA token & Ricart-Agrawala's token-based mutual exclusion~\cite{RA83}
  & 67 &   8 &   8 &     N/A &   N/A &     967 &  31.30 \\\hline
  SK token & Suzuki-Kasami's token-based mutual exclusion~\cite{SK85}
  & 72 &   6 &  11 &     N/A &   N/A &     760 &  22.75 \\\hline
  CR leader & Chang-Robert's leader election~\cite{CR79}
  & 30 &   2 &   1 &      81 &  1.34 &     114 &   1.69 \\\hline
  HS leader & Hirschberg-Sinclair's leader election~\cite{HS80}
  & 51 &   2 &   2 &     135 &  1.84 &     170 &   2.23 \\\hline
  2P commit & Two-phase commit~\cite{Gray78}
  & 82 &  16 &   6 &    1137 & 32.20 &    1345 &  43.11 \\\hline
  DS crash & Dolev-Strong's consensus under crash failures~\cite{DS83}
  & 52 &   6 &   8 &     N/A &   N/A &     626 &  17.53 \\\hline
  CL Paxos & Castro-Liskov's Paxos under Byzantine failures~\cite{CL02byz}
  &174 &  22 &  14 &    1517 & 64.97 &    1624 &  74.47 \\\hline
  \hline

  Birthday & Birthday Collision~\cite{blog16examp-gsite}
  & 14 &   4 &   4 &     173 &  4.77 &     201 &   5.55 \\\hline
  Cite author & Publication Citation Matching author query~\cite{blog16examp-gsite}
  & 13 &   2 &   3 &      59 &  1.47 &      87 &   1.92 \\\hline
  Cite exist & Publication Citation Matching existential query~\cite{blog16examp-gsite}
  & 19 &   2 &   3 &      65 &  1.81 &      93 &   2.12 \\\hline
\end{tabular}\smallskip\\
\begin{tabular}{@{}p{\textwidth}@{}}
  For \c{CheckAccess} in Core RBAC, the original LOC is for the entire
  Core RBAC, and only queries needed for the \c{CheckAccess} function
  are incrementalized. 

  For distributed algorithms, the original LOC is for queries and updates
  after quantifications are transformed into aggregate queries and nested
  queries using the method in~\cite{Liu+12DistPL-OOPSLA} and its interface
  with our system.
  Two of the algorithms are not listed here due to incomplete interface in
  transforming witness for existential quantifications.
  The N/A entries for three of the algorithms are due to incomplete interface
  for transforming set values sent in messages; 
  this does not affect generation of filtered programs because demand-driven 
  queries can be computed after the complete set value is received.

  For probabilistic queries, the original LOC is also for queries and updates
  after transforming quantifications into aggregate and nested queries.
\end{tabular}

\caption{Program and transformation statistics: 
  size of original, incremental, and filtered programs in lines of code (LOC);
  number of queries incrementalized and of updates to those queries;
  and transformation time to generate the incremental and filtered programs, 
  respectively. %
}
  \label{tab-all}
\end{table*}

\section{Related work and conclusion}
\label{sec-relate}

There is much previous work on incremental
computation~\cite{RamRep93,Liu13book}, including demand-driven
incremental computation, for lower-level programs, high-level
relational queries, and object queries.

There are numerous methods for incremental computation in lower-level
languages: memo functions~\cite{Michie68}, caching in functional
programs~\cite{Pugh:Teitelbaum:89}, static incrementalization of recursive
functions and loops~\cite{Liu00IncEff-HOSC}, memoization and dynamic
dependence graphs for functional and imperative programs~\cite{Acar09},
function caching in object-oriented programs~\cite{shankar07ditto}, and
making dependence-graph-based approaches demand-driven and
composable~\cite{Hammer14}, among others.
These methods do not handle high-level queries under all updates that
affect the query results

There has been significant work on incremental computation of high-level
queries over sets and relations, in set languages,
relational databases, and logic languages,
e.g.,~\cite{GupMum99maint}: %
using finite differencing rules to incrementalize set
expressions~\cite{PaiKoe82}, %
deriving production rules for incremental view
maintenance~\cite{CerWid91}, %
incremental computation of relational algebra
expressions~\cite{QiaWie91,GriLibTri97}, %
incremental view maintenance for recursive views~\cite{GupMumSub93},
Datalog queries with complexity
guarantees~\cite{LiuSto09Rules-TOPLAS},
demand-driven computation using magic sets (MST)~\cite{Ban+86}, demand
transformation (DT) with complexity
guarantees~\cite{TekLiu11RuleQueryBeat-SIGMOD}, %
scale-independent relational queries~\cite{Arm+13},
implementation in newer languages~\cite{mit2014i3ql}, and incremental
view maintenance for nested queries~\cite{lupei2014nested},
among others.
These methods do not handle arbitrarily nested and aliased sets and
objects.

Filtering %
using demand is similar to the idea underlying MST~\cite{Ban+86} and
DT~\cite{%
  TekLiu11RuleQueryBeat-SIGMOD}.  The difference, besides handling
arbitrary sets and objects, is that our tag sets and filtered relations are
asymptotically no larger than the given data, whereas MST and DT may use
asymptotically larger space.
For example, MST and DT may need \O{n_1\times n_2} space to store the join
of two %
relations of sizes \O{n_1} and \O{n_2}, whereas our method requires at most
\O{n_1+n_2} space.
The trade-off is that this stored join may help reduce the asymptotic query
time more than our method does.

Incremental computation of object queries has been studied in object
databases and object-oriented programming
languages~\cite{SalMez13}\full{}.
We discuss closest related work below.  
Note that our method allows powerful object queries to be written
completely declaratively, contrasting previous work, as discussed in
Section~\ref{sec-lang}.

Caching the results of query functions in object databases,
e.g.,~\cite{Kemper91odb-caching}, allows the results to be reused when
queries on the same keys are encountered again\full{}.  The problem with
caching for queries over sets of objects is that, unless all objects
and sets used are taken as keys for cache lookup, which is not
feasible in general,
the cached results may become invalid and must be discarded when any
of the values not taken as keys are updated.  In contrast, our method
maintains the cached results incrementally as values used in the
queries are updated.

Incremental view maintenance using update propagation, e.g.,
\cite{KunRun98,Nak01}, dynamically tracks and propagates changes.  A
particularly clean method~\cite{Nak01} was implemented using
Smalltalk: it translates all values
into collections, including scalar values into singleton sets, and
object values into sets of attribute name-value pairs,
and handles all collections and updates uniformly; it gives no cost
analysis, although some other work does~\cite{KunRun98}.
However, these methods do not define invariants for propagating demand
and change or for creating indices, making the formulation of those
complex mechanisms ad hoc.  These dynamic methods also do not generate
incremental maintenance code, leading to the unnecessary overhead and
hard-to-predict behavior characteristic of dynamic methods.

Liu et al.~\cite{Liu+05OptOOP-OOPSLA} studies incrementalization for
object-oriented programs.  It transforms expensive queries and updates
by applying manually written incrementalization rules from a library
of rules.
How to automatically generate the incremental maintenance code needed
in such rules was left open; it was later studied for relational
queries, in implementing core RBAC~\cite{Liu+06ImplCRBAC-PEPM}, but
that method does not handle nested and aliased sets and objects.
Our method translates nested and aliased sets and objects into flat
field and member relations as in~\cite{RotLiu08OSQ-GPCE}, not into
nested sets with attribute name-value pairs for objects as
in~\cite{Nak01}, because the former allows simpler and more efficient
implementations, essentially as in column databases~\cite{Stone05c}.

Rothamel and Liu~\cite{RotLiu08OSQ-GPCE} generates code for
incremental maintenance of query results for arbitrary objects and
sets, but tracks demand and maintains indices using dynamic
mechanisms, and propagates change by dynamically assigning and
invoking maintenance code as obligations on objects.  The complex
dynamic mechanisms
made it exceedingly difficult to understand the method, let alone
predict its performance and improve it.
Our %
method of defining everything using invariants allowed us to generate
complete code, provide precise complexity guarantees,\short{ and}
develop a refined demand mechanism that gives asymptotic improvement
over~\cite{RotLiu08OSQ-GPCE}, as shown in
Section~\ref{sec-expe-osq}\full{}.

JQL extends Java with object queries~\cite{Willis06,Willis08}.  The
class of queries and updates it incrementalizes is very restricted:
only queries over source sets of objects with tests against fields of
these objects, and only updates to the source sets and fields of the
element objects.
Implemented using AspectJ, the method used
for join queries
requires iterating over entire sets, not just changes determined using
incrementally maintained indices as in our method, and thus yields
asymptotically slower programs than our method, as shown in
Section~\ref{sec-expe-jql}.
A recent work
builds on JQL by adding more sophisticated query planning and by caching
intermediate results using various
strategies~\cite{nerella2014exploring,nerella2014efficient}; it shows
constant-factor improvements but has
the same lack of auxiliary indices that leads to asymptotic slowdowns.

In conclusion, by establishing invariants for not only query results
but also auxiliary values for tracking demand and propagating change,
our method can generate complete
implementations of demand-driven incremental queries with precise
complexity guarantees.
Additional details and extensions can be found in~\cite{Bra16thesis}.
Future work includes more efficient range queries,
optimal selection of specialized incremental maintenance when appropriate,
improved static analysis for reducing constant factors,
parallel computation for independent maintenance of query results,
and asynchronous maintenance of auxiliary indices.

\myparag{\large Acknowledgment}
We thank \anony{Anonymized}{Xuetian Weng} for help making the JQL
implementation work and for experiments using the Zipf distribution.

\def\bibdir{../../../bib}                  %
{%
\bibliography{\bibdir/strings,\bibdir/liu,\bibdir/IC,\bibdir/PT,\bibdir/PA,\bibdir/Lang,\bibdir/Algo,\bibdir/Perform,\bibdir/DB,\bibdir/SE,\bibdir/Sys,\bibdir/Sec,\bibdir/misc,\bibdir/crossref}
\bibliographystyle{abbrvnat}
}

\full{} %

\appendix

\section{Example for queries with multiple occurrences of a %
  relation}
\label{sec-self-join}

For the running example, \M{} occurs twice in the relational query,
and thus there are two maintenance blocks for an update to \M:
\begin{code}
// maintenance block for \M{} += \{(fs,user)\}
// corresponding to clause \c{(fs,user) in \M}
if fs in \T{fs} and user in \T{user}:
  for e in \img{\F{email}}{user}:
    for l in \img{\F{loc}}{user}:
      if cond(l):
        for celeb in \img{\S{\inv{\F{followers}}}}{fs}:
                         // \#\img{\S{\inv{\F{followers}}}}{fs}
          for group in \img{\U.2}{1=celeb}:
                         // \#\img{\U.2}{1=celeb}
            if user in \img{\M}{group}:
              r add (celeb,group,e)
\end{code}%
\begin{code}
// maintenance block for \M{} += \{(group,user)\}
// corresponding to clause \c{(group,user) in \M}
if group in \T{group} and user in \T{user}
  for e in \img{\F{email}}{user}:
    for l in \img{\F{loc}}{user}:
      if cond(l):
        for celeb in \img{\U.1}{2=group}:
                         // \#\img{\U.1}{2=group}
          for fs in \img{\F{follower}}{celeb}:
            if user in \img{\M}{fs}:      
              r add (celeb,group,e)       
\end{code}
For the first block, the filtered auxiliary values used are defined as
follows; maintaining them at updates to \M{} requires no work.
\begin{code}
   \T{celeb} \= \{celeb: (celeb,group) in U\}
   \S{\inv{\F{followers}}}
   \= \{(fs,celeb): (celeb,fs) in \F{followers},
                     celeb in \T{celeb}\}
\end{code}
Concatenating the two blocks and augmenting the second block with a
test, to exclude the element added for the first occurrence of \M{} in
the query, yield the following maintenance for the query result for
update \c{\M{} += \{(s,u)\}}:
{\label{code-selfjoin}
\begin{code}
// maintenance block for \M{} += \{(s,u)\}
// corresponding to clause \c{(fs,user) in \M}
if s in \T{fs} and u in \T{user}:   // fs -> s, user -> u
  for e in \img{\F{email}}{u}:                   // user -> u
    for l in \img{\F{loc}}{u}:                   // user -> u
      if cond(l):
        for celeb in \img{\S{\inv{\F{followers}}}}{s}: // fs->s
          for group in \img{\U.2}{1=celeb}:
            if u in \img{\M}{group}:           // user -> u
              r add (celeb,group,e)
// maintenance block for \M{} += \{(s,u)\} 
// corresponding to clause \c{(group,user) in \M},
// with augmentation for clause \c{(fs,user) in \M}
if s in \T{group} and u in \T{user}: // group->s, user->u
  for e in \img{\F{email}}{u}:                   // user -> u
    for l in \img{\F{loc}}{u}:                   // user -> u
      if cond(l):
        for celeb in \img{\U.1}{2=s}:          // group-> s
          for fs in \img{\F{follower}}{celeb}:
            if u in \img{\M}{fs}:              // user -> u
              if (fs,u) != (s,u)     // added test
                r add (celeb,s,e)            // group-> s
\end{code}\full{}}

\section{Transformations back to implementations on objects}
\label{sec-phase3}

\short{
Table~\ref{tab-back} presents the transformations.
\begin{table}[htb]
\begin{center}
\small
\begin{tabular}{@{}l@{\,}|@{~}l@{\hspace{0ex}}}
  \hline
  operations on relations     & generated operations on objects\\\hline\hline

  \c{if (x,y) in \F{f}:}      & \c{if x hasfield f:}\\
  \c{~~stmt}                  & \c{~~if y == x.f:}\\
                              & \c{~~~~stmt}\\\hline
  \c{for y in \img{\F{f}}{x}:}& \c{if x hasfield f:}\\
  \c{~~stmt}                  & \c{~~y = x.f}\\
                              & \c{~~stmt}\\\hline\hline

  \c{if (x,y) in \M}:         & \c{if x isset:}\\
  \c{~~stmt}                  & \c{~~if y in x:}\\
                              & \c{~~~~stmt}\\\hline
  \c{for y in \img{\M}{x}:}   & \c{if x isset:}\\
  \c{~~stmt}                  & \c{~~for y in s:}\\
                              & \c{~~~~stmt}\\\hline\hline

  \c{if xy in \imge{\U.Ixy}:}     & \c{if xy in \U{}Ixy:}\\
  \c{~~stmt}                      & \c{~~stmt}\\\hline
  \c{for y in \img{\U.Iy}{Ix=x}:} & \c{if x in \dom{\U{}IxIy}:}\\
  \c{~~stmt}                      & \c{~~for y in \imgo{\U{}IxIy}{x}:}\\
                                  & \c{~~~~stmt}\\\hline
  \c{for x in \img{\U.Ix}{Iy=y}:} & \c{if y in \dom{\U{}IyIx}}:\\
  \c{~~stmt}                      & \c{~~for x in \imgo{\U{}IyIx}{y}:}\\
                                  & \c{~~~~stmt}\\\hline
  \c{for xy in \imge{\U.Ixy}:}    & \c{for xy in \U{}Ixy:}\\
  \c{~~stmt}                      & \c{~~stmt}\\\hline\hline

  \c{for x in \img{r}{y}:}        & \c{if y in \dom{r}:}\\
  \c{~~stmt}                      & \c{~~for x in \imgo{r}{y}:}\\
  for filtered \c{r}              & \c{~~~~stmt}\\\hline\hline

  \c{s add x}, for tag set \c{s} & \c{s.cadd(x)}\\\hline
  \c{s del x}, for tag set \c{s} & \c{s.cdel(x)}\\\hline\hline
  
  \c{r += \{(x,y)\}}, for filtered \c{r} & \c{r.add(x,y)}\\\hline
  \c{r -= \{(x,y)\}}, for filtered \c{r} & \c{r.del(x,y)}\\\hline\hline

  \c{r add (p1,...,pk,x)}, for result \c{r} & \c{r.cadd((p1,...,pk),x)}\\\hline
  \c{r del (p1,...,pk,x)}, for result \c{r} & \c{r.cdel((p1,...,pk),x)}\\\hline
  \img{r}{(p11,...,pk)}, for result \c{r}   & \imgo{r}{(p1,...,pk)}\\
                                            & \c{~~~if (p1,...,pk) in \dom{r}}\\
                                            & \c{\{\} otherwise}\\
  \hline
\end{tabular}
\end{center}
\todo{}
\c{xy} denotes a tuple with components from both \c{x} and \c{y}, 
subsets of components of \U.~
\c{Ix}, \c{Iy}, and \c{Ixy} denote the indices  corresponding to 
components in \c{x}, \c{y}, and \c{xy}, respectively.~
\c{\U{}Ixy} is the set of tuples of components \c{Ixy} in \U.~
\c{\U{}IxIy} is the map mapping components \c{Ix} to components \c{Iy}
in \U.~  \c{\U{}IyIx} is symmetric with \c{\U{}IxIy}.  
  \caption{Transformations to operations on objects.}
  \label{tab-back}
\end{table}
  Note that:
\begin{itemize}

\item There are no entries for \c{for (x,y) in \F{f}} and
  \c{for}\linebreak\c{(x,y) in \M}, because they are slower than the
  corresponding code used in other orders of retrieving matching values and
  hence never appear in our generated operations on relations.

\item There are not entries for \c{if y in \T{x}}; they do not need to be
  transformed.

\item There are no entries for updates to \F{f} and \M; they are exactly
  the opposite of Phase 1 transformations for updates.

\item There are no entries for updates to \U; they do not need to be
  transformed.

\item In the two entries that use \c{\U{}Ixy}, to reduce space, uses
  of \c{\U{}Ixy} can be replaced with uses of \c{\U{}IxIy} or
  \c{\U{}IyIx} if either is already used.

\end{itemize}
} %

\section{Effect of demand set size}
\label{sec-demand-set-size}

Filtering using demand provides the most benefit when the demand set
is small relative to the entire domain, due to the overhead of
maintaining tag sets and filtered relations.  To evaluate the effect
of \c{\#\U}, we ran experiments to compare incremental and filtered
programs for different \c{\#\U}.

We use the setup for 20,000 users as in the rightmost input for
Figure~\ref{fig-celeb-asymp}, except with the number of users in \U{}
ranging from 1 to all 20,000.  %
Also, we do location updates to only users in the tag set for
\c{user}; otherwise, the filtered program always gains significantly
by skipping maintenance for location updates to other users.

\begin{figure}[htb]
  \centering
  \vspace{-2ex}

  \hspace{-2ex}
  \includegraphics[width=3in]{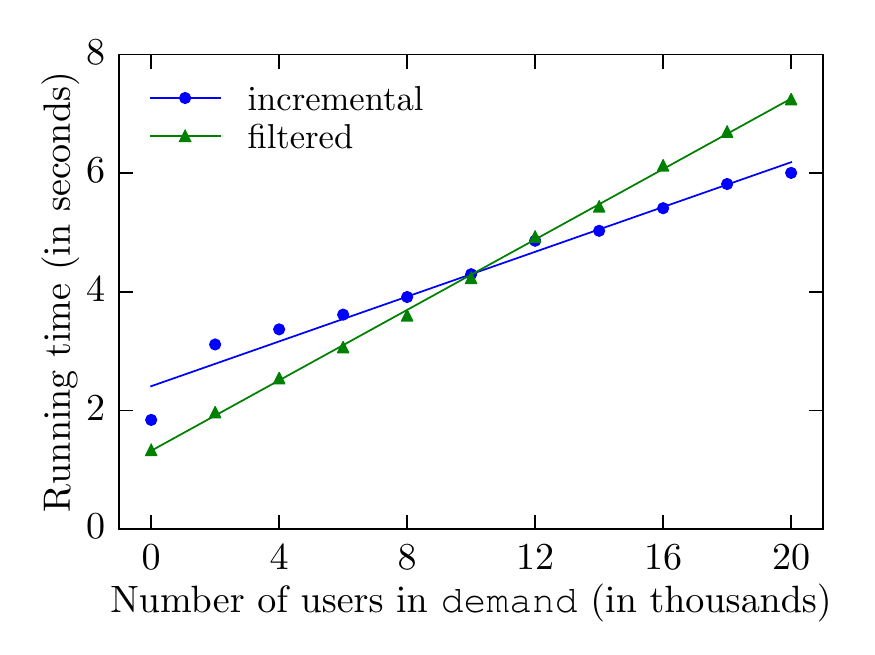}\vspace{-2.5ex}

  \hspace{-2ex}
  \includegraphics[width=3in]{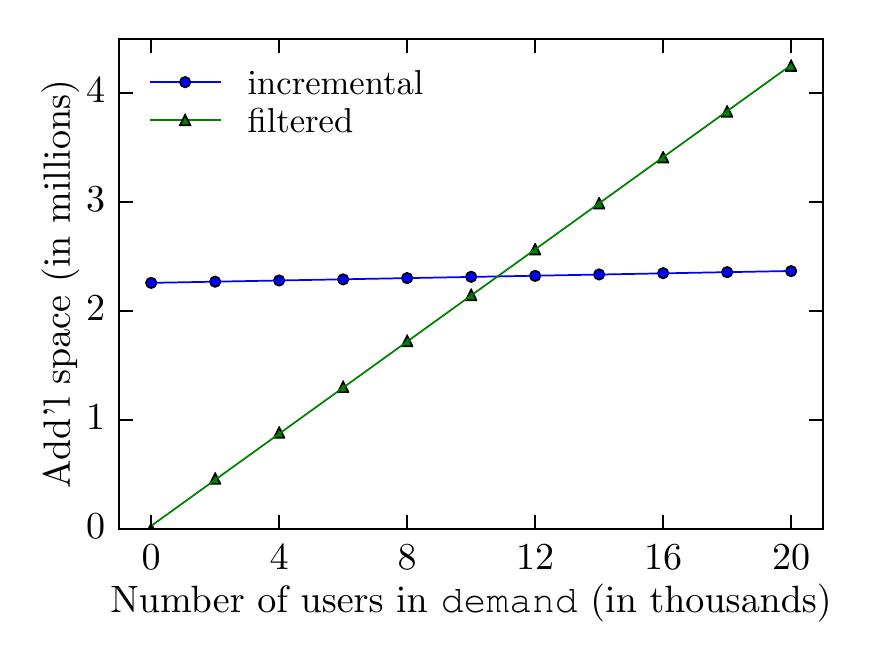}\vspace{-2.5ex}

\caption{Running time and additional space for varying demand set size.}
\label{fig-celeb-demand}
\end{figure}

Figure~\ref{fig-celeb-demand}-top and -bottom show the measured time
and additional space, respectively, for 200,000 repeats of a
query-update pair as for Figure~\ref{fig-celeb-asymp}.
\begin{itemize}

\item For running time, the increase for the incremental program, 
  is due to increase in expected \c{\#\img{\U.2}{1=celeb}} from nearly 0 to
  1 in running the maintenance inside the loop over
  \c{\img{\U.2}{1=celeb}};
  the faster increase for the filtered program is due to the linear
  increase in \c{\#\img{\S{\inv{\M}}}{user}}.

\item For additional space, the main usage for the incremental program
  is for auxiliary relations, and the slight increase is due to
  increase in stored query result sets;
  the steady increase for the filtered program is due to increase in
  size of tag sets and filtered relations.

\end{itemize}
Clearly, the filtered program is faster and uses less additional space
when the ratio of users in \U{} to all users is small, i.e. when
\c{\#\U} is relatively small.
In general, the crossover points depend on the query, updates, and
data, but the gain is obvious for ratios below 10\% and is small or
negative for ratios above 50\%.

\full{} %

\end{document}